\documentclass[sigconf]{acmart} 

\AtBeginDocument{%
  \providecommand\BibTeX{{%
    \normalfont B\kern-0.5em{\scshape i\kern-0.25em b}\kern-0.8em\TeX}}}

\setcopyright{none}

\newcommand{\planninglogo}{\includegraphics[scale=0.2]{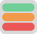}}

\newcommand{\tool}{\textsc{Rehearsal}}
\usepackage{wrapfig}
\usepackage{booktabs} 
\newcommand{\rev}[1]{{{#1}}} 

\copyrightyear{2024}
\acmYear{2024}
\setcopyright{rightsretained}
\acmConference[CHI '24]{Proceedings of the CHI Conference on Human Factors
in Computing Systems}{May 11--16, 2024}{Honolulu, HI, USA}
\acmBooktitle{Proceedings of the CHI Conference on Human Factors in
Computing Systems (CHI '24), May 11--16, 2024, Honolulu, HI, USA}
\acmDOI{10.1145/3613904.3642159}
\acmISBN{979-8-4007-0330-0/24/05}

\begin{document}

\title{Rehearsal: Simulating Conflict to Teach Conflict Resolution}


\author{Omar Shaikh}
\affiliation{%
  \institution{Stanford University}
  \city{Stanford}
  \country{USA}}
\email{oshaikh@stanford.edu}

\author{Valentino Chai}
\affiliation{%
  \institution{Stanford University}
  \city{Stanford}
  \country{USA}}
\email{vechai@stanford.edu}

\author{Michele J. Gelfand}
\affiliation{%
  \institution{Stanford University}
  \city{Stanford}
  \country{USA}}
\email{gelfand1@stanford.edu}

\author{Diyi Yang}
\authornote{Both authors co-advised.}
\affiliation{%
  \institution{Stanford University}
  \city{Stanford}
  \country{USA}}
\email{diyiy@stanford.edu}

\author{Michael S. Bernstein}
\authornotemark[1]
\affiliation{%
  \institution{Stanford University}
  \city{Stanford}
  \country{USA}}
\email{msb@stanford.edu}

\renewcommand{\shortauthors}{Shaikh, et al.}

\begin{abstract}
    Interpersonal conflict is an uncomfortable but unavoidable fact of life. Navigating conflict successfully is a skill---one that can be learned through deliberate practice---but few have access to effective training or feedback. To expand this access, we introduce \tool{}, a system that allows users to rehearse conflicts with a believable simulated interlocutor, explore counterfactual ``what if?'' scenarios to identify alternative conversational paths, and learn through feedback on how and when to apply specific conflict strategies. Users can utilize \tool{} to practice handling a variety of predefined conflict scenarios, from office disputes to relationship issues, or they can choose to create their own setting. To enable \tool{}, we develop \textit{IRP prompting}, a method of conditioning output of a large language model on the influential Interest-Rights-Power (IRP) theory from conflict resolution. \tool{} uses IRP to generate utterances grounded in conflict resolution theory, guiding users towards counterfactual conflict resolution strategies that help de-escalate difficult conversations. In a between-subjects evaluation, $40$ participants engaged in an actual conflict with a confederate after training. Compared to a control group with lecture material covering the same IRP theory, participants with simulated training from \tool{} significantly improved their performance in the unaided conflict: they reduced their use of escalating competitive strategies by an average of 67\%, while \textit{doubling} their use of cooperative strategies. Overall, \tool{} highlights the potential effectiveness of language models as tools for learning and practicing interpersonal skills.
\end{abstract}




\maketitle


\section{Introduction}
Managing interpersonal conflict is a critical skill. We occasionally find ourselves in situations where our interests, values, or goals conflict with others. If left unchecked, conflict can reach a boiling point, manifesting in verbal arguments, physical altercations, passive-aggressive behavior, or more~\cite{burton1990conflict, de2004conflict}. Additionally, conflict correlates with increased stress \cite{friedman2000goes}, a downturn in productivity, and absenteeism \cite{jehn2006effects}. While avoiding any conflict may be impractical~\cite{marx_engels_1848}, how we choose to deal with conflict is not: in most settings, an ideal outcome for both parties is to work \textit{cooperatively}~\cite{lytle1999strategic}. 

Directing conflict towards cooperative communication is, however, a difficult skill to learn, requiring targeted and repeated practice with immediate feedback~\cite{deutsch1994constructive}. Avenues for practicing conflict resolution are unfortunately often limited: training material for conflict resolution is usually static (e.g. a written case study) covering a fixed number of situations. Independently extrapolating beyond these predefined settings---especially without expert guidance---is challenging. While conflict roleplay with an expert is a proven and widely used technique~\cite{ebner2005using}, expert training is costly and scarce. If it were possible to simulate expert-level conflict practice, we could significantly improve an individual's conflict resolution skills in a cost-effective and scalable manner.

We envision that, given their generative capabilities~\cite{brown2020language}, large language models (LLMs) offer an opportunity to craft expert-level conflict roleplays and provide immediate feedback to users. Despite remarkable progress in producing compelling content, however, LLMs such as ChatGPT often fall short of simulating conflict and giving feedback on it. Naively prompting LLMs introduces a host of problems that lead to unrealistic and ineffective simulations. \textbf{First, current LLMs are sycophantic} due to instruction following, producing generations that agree too quickly with the viewpoints of a user \cite{perez2022discovering}. \textbf{Second, providing targeted practice and feedback is challenging due to the open-endedness of LLM text generation.} An off-the-shelf LLM may produce messages that are not directly informative---potentially even distracting---for teaching conflict resolution. In contrast, students benefit significantly from deliberate and targeted practice \cite{ericsson1993role}, where feedback is readily available \cite{hattie2007power, schwartz2016abcs}. As a result, experts teaching conflict resolution rely on a proven and curated set of conflict resolution strategies when roleplaying conflict, providing feedback specific to each strategy~\cite{lytle1999strategic}. Applying these strategies without guidance and in a fully open-ended setting can be a significant burden for students. If a student unintentionally uses an inappropriate conflict resolution strategy, they can send a conversation into a conflict spiral~\cite{brett1998breaking}. 

\begin{figure}
  \begin{center}
    \includegraphics[width=0.45\textwidth]{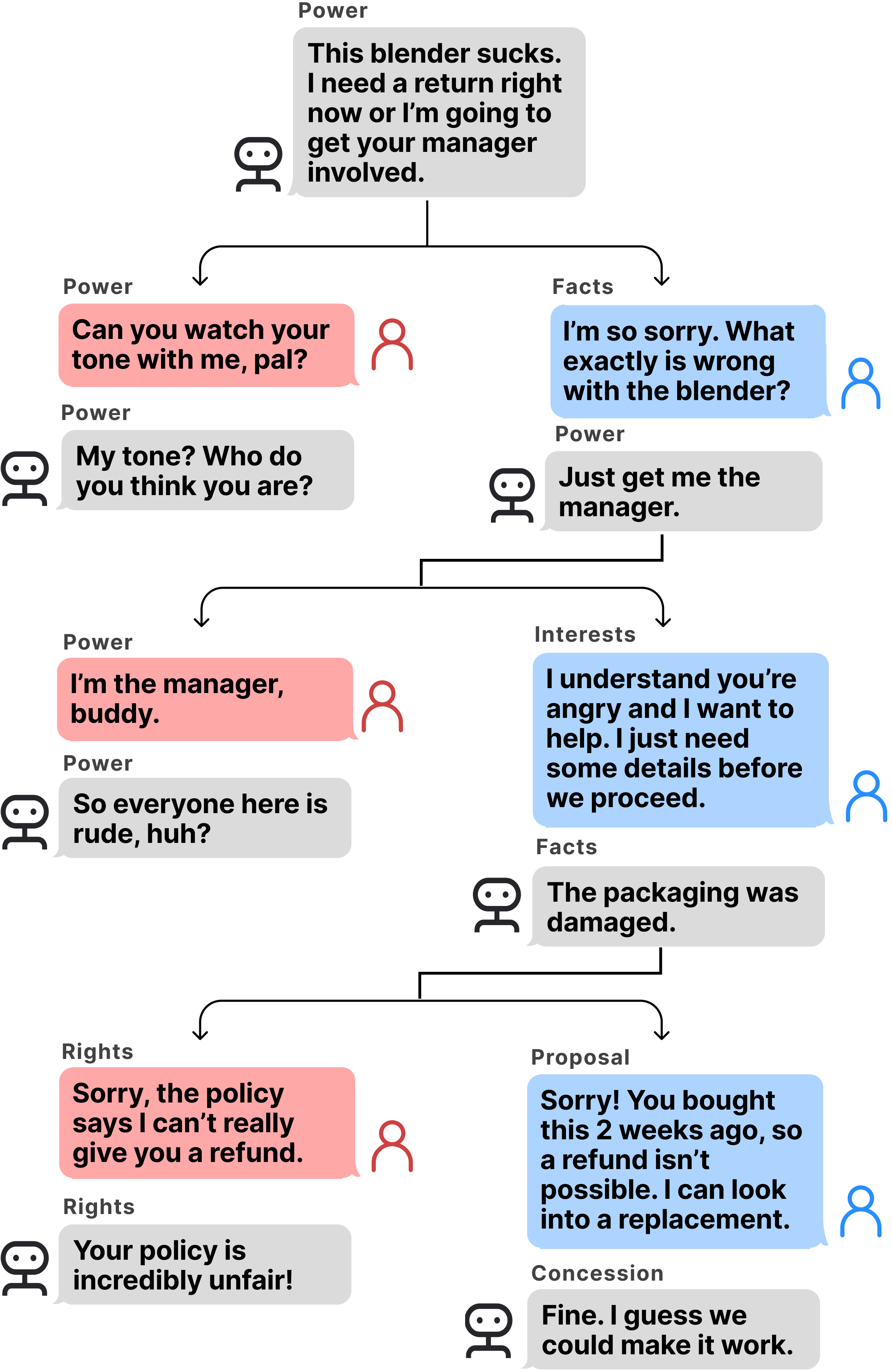}
  \end{center}
  \caption{An example interaction trace with \tool{}, where an employee practices conflict with a simulated customer. The employee quickly realizes that Rights and Power-based strategies result in heightened conflict. The conflict is eventually resolved using an Interests-oriented approach.}
  \label{fig:interaction_trace}
\end{figure}

To mitigate challenges associated with simulating conflict, we turn to conflict resolution theory as grounding for LLMs. At its core, effective conflict resolution relies on interest-based discussion of conflict. We draw on the \textit{Interests-Rights-Power (IRP)} framework from the conflict literature~\cite{lytle1999strategic} to ground our approach. The IRP framework codes utterances from individuals onto a higher-level set of 8 strategies. For example, a \textit{Power} strategy threatens to apply consequences to the other party: ``I'll be telling everyone about what you did here today.'' \textit{Rights} asserts a standard of fairness or legitimacy: ``I was on call yesterday, and we rotate the job.'' \textit{Interests} discuss or draw on each party's independent goals: ``If we do this together and win, it will show up on your record for promotion next month.'' 
According to IRP, individuals in a conflict should discuss each party's \textit{Interests} without resorting to threats (\textit{Power}) or deferring to pre-existing norms (\textit{Rights})~\cite{ury1988getting}. In a state of conflict, however, we are not often aware of how the strategies we use will cascade as the conflict proceeds. For example, deploying a Power strategy tends to result in the other party responding with Power as well---and when both parties repeatedly use Power or Rights strategies, the conflict is unlikely to end productively~\cite{ury1988getting}.\footnote{~``I'm pulling the plug on your CHI submission---it's not ready.'' [Power strategy]\\``NO! What a betrayal! I've worked on this project nonstop all year, and I need this published to be competitive on the job market!'' [Rights strategy]\\``I'm sorry, but I'm your advisor and senior author on the paper. I make the call here.'' [Rights strategy]\\``Not when I tell all the new Ph.D. students about how you kneecap your students.'' [Power strategy]\\-- illustrative example; don't worry, the coauthors of this paper all get along just fine.} To avoid conflict spirals, one should (if possible) rely primarily on cooperative strategies like Interests and avoid contentious strategies like Rights or Power~\cite{brett1998breaking}. Experts rely heavily on frameworks like IRP to jointly provide feedback and offer targeted roleplays of specific conflict scenarios~\cite{gill2015inside}.

Using these insights, we operationalize the IRP framework as a basis for simulating teachable conflict and enabling expert feedback with LLMs. We introduce \textit{IRP Prompting}, a method for steering LLM generations with a theoretical conflict resolution framework. In IRP Prompting, the IRP framework plays a core role in planning the course of a conflict, serving as grounding for the LLM. IRP prompting builds on multi-step prompting techniques~\cite{wei2022chain, yang-etal-2022-re3, wu2022ai}: instead of directly generating conflict dialogue---sampling from an exponentially large space of potentially undesirable outputs---we first classify the simulated interlocutor's next conflict strategy based on the current conversation, then generate messages conditioned on that explicit conflict resolution strategy. Compared to out-of-the-box generations from an LLM, we find that IRP prompting produces conflict simulations that are more representative of expert roleplay. Grounding in IRP also yields simulations robust to repeated interaction with a user: simulations can only reach a state of agreement after multiple interest-based strategies are used. Tightly integrating conflict resolution theory with LLMs \textbf{enables targeted feedback} on the use of conflict resolution strategies and \textbf{mitigates teaching limitations} with open-ended conflict generation. 

We instantiate IRP prompting in an interactive dialogue system, \textbf{\tool{}}, where users engage directly with a simulated interlocutor. Using \tool{}, users receive targeted feedback at each dialogue turn and can explore how a simulated interlocutor might respond if different strategies were employed. For example, consider a new employee who wants to practice managing conflict with a dissatisfied customer (Figure \ref{fig:interaction_trace}). The employee can specify a conflict setting from their own life or pick a predefined one from \tool{} for practice. The simulated interlocutor starts by using \textit{Power}, threatening to involve management. Taken aback, the employee responds with Power, tone-policing the simulated customer. Here, \tool{} (using IRP prompting) generates several alternate messages that use more cooperative strategies and scores each alternative. Looking at the generated alternatives, the employee decides to use Interests, identifying why the customer was angry. After a few interactions, the employee realizes that using Rights (\textit{``we don't do returns here''}) or Power (\textit{``get out of my store!''}) early in a conversation results in heightened conflict. By observing alternative strategies across multiple interactions, the employee learns that an Interests-based approach---brainstorming a potential compromise with the customer---yields significantly better outcomes. 

We conduct evaluations across two axes: a technical analysis of IRP prompting and a behavioral study of \tool{}'s end-to-end effectiveness as a system for teaching conflict resolution. First, we compare IRP prompting to out-of-the-box LLMs and to ablations, evaluating generations in a controlled setting. Compared to our ablations, we find that IRP prompting produces conflict simulations more faithful to expert training. Next, we recruit $N=40$ participants and study \tool{}'s effectiveness by comparing it with traditional conflict resolution training. In a between-subjects evaluation, participants engaged in an actual conflict following training with or without \tool{}. Participants with \tool{} training significantly improved their application of effective conflict resolution strategies in the unaided conflict, despite not showing differences in ``book smarts'' recognition or recall on a test. Compared to a control group, \tool{} participants reduced their use of competitive strategies by an average of 67\% while \textit{doubling} their use of cooperative strategies.

\begin{table*}[]
    \centering
    \begin{tabular}{l|p{6.5cm}p{4.5cm}}
        \toprule
        \textbf{Strategy} & \textbf{Definition} & \textbf{Example}  \\
        \midrule
        \multicolumn{3}{c}{\textit{Cooperative Strategies}} \\
        \midrule
        Interests & Reference to the wants, needs, or concerns of one or both parties. This may include questions about why the negotiator wants or feels the way they do. & \textit{We can figure this out---I understand that you've been really busy lately.} \\
        \midrule
        Positive Expectations & Communicating positive expectations through the recognition of similarities and common goals & \textit{I know you're an excellent employee and I want to make sure you get a promotion.} \\
        \midrule
        Proposal & Proposing concrete recommendations that may help resolve the conflict & \textit{Why don't we record your progress weekly instead of monthly, so we can stay on track?} \\
        \midrule
        Concession & Changing an initial view or position (in response to a proposal) to resolve a conflict & \textit{That makes sense---I'll try recording my weekly progress instead of doing it monthly.} \\
        \midrule 
        \multicolumn{3}{c}{\textit{Neutral Strategies}} \\
        \midrule
        Facts & Providing information on the situation or history of the dispute, including requests for information, clarification, or summaries. & \textit{Unfortunately, I haven't been able to keep track of your progress over the last several weeks.} \\
        \midrule 
        Procedural & Introductory messages, including discussion about discussion topics, procedures, etc. & \textit{Hi! How are you? Do you have time today to talk about a promotion?} \\
        \midrule         
        \multicolumn{3}{c}{\textit{Competitive Strategies}} \\
        \midrule
        Power & Using threats and coercion to try to force the conversation into a resolution. & \textit{I'm going to tell everyone you've been missing deadlines.} \\
        \midrule
        Rights & Appealing to fixed norms and standards to guide a resolution. & \textit{Sorry, I can't do anything---company policy doesn't allow that.} \\
        \bottomrule
    \end{tabular}
    \caption{A set of conflict resolution strategies and corresponding examples from \citet{brett1998breaking}. Strategies can be separated into three broad categories: \textit{cooperative}, \textit{neutral}, and \textit{competitive}. To ground LLM generations, we strategically plan and generate messages that adhere to the above strategies.}
    \label{tab:conflict_strategies}
\end{table*}

In summary, \tool{} highlights the potential of generative AI as a teaching tool for social interaction skills, combining social scientific research on conflict resolution theory~\cite{fisher2011getting} with research on the generative capabilities of LLMs~\cite{brown2020language}. We contribute:
\begin{enumerate}
    \item \tool{}: an interactive system for roleplaying conflict resolution. In a simulated conflict roleplay, \tool{} generates feedback and lets people send/evaluate their messages in a roleplay. Furthermore, \tool{} enables learning from alternative conflict resolution strategies.  

    \item \textit{IRP Prompting}: a prompting technique for producing conflict faithful to expert training by grounding LLM generations to conflict resolution theory. IRP prompting also supports generating alternative messages (that use a different conflict resolution strategy), enabling \tool{}'s interactions.

    \item An evaluation of \textit{IRP prompting} and a user study of \tool{} with $N=40$ participants. Our studies highlight \tool{}'s significant effectiveness in applying conflict resolution strategies, compared to the status quo of teaching the same material. 
\end{enumerate}

\section{Related Work}
To design and build \tool{}, we rely heavily on research from conflict resolution theory, large language models, and computer-supported cooperation.

\subsection{Dealing with Interpersonal Conflict}
\label{conflict_theory}
Broadly, conflict resolution seeks to facilitate the end of conflict and provides frameworks to convert conflicting situations into cooperative ones. Conflict resolution theory splits types of conflict into two states: cooperative and competitive \cite{deutsch1949theory}. Ideally, conflict should move towards a cooperative state, where all parties discuss common goals and interests. This is harder said than done---the underlying processes behind effective conflict resolution are incredibly delicate. Concrete conflict resolution \textit{strategies} provide pathways to reaching a cooperative state. Strategies are split into two broad categories: constructive and destructive \cite{deutsch1973resolution}. Constructive strategies seek to cooperatively integrative perspectives from both parties, while destructive strategies do the opposite.

While specific resolution strategies can be broadly construed as constructive or destructive, effectively using them in practice requires a narrower set of definitions. To this end, more recent work has expanded on \citet{deutsch1973resolution}'s initial resolution work. Instead of framing conflict as constructive vs. destructive, \citet{ury1988getting} frames conflict as a division across three categories: \textit{Interests}, \textit{Rights}, and \textit{Power}. At a high level, the \textit{Interests} strategy focuses on building common ground between both participants. While using \textit{Interests}, both parties will actively problem solve, cooperatively discovering interests that lead to an eventual resolution. Importantly, \textit{Interests} focuses on healthy future outcomes of a conflict, integrating concerns, needs, fears, and desires of both parties (e.g. \textit{``Let's try to work things out here''}). Cooperative conflict builds mostly on interests. However, conflict usually escalates to \textit{rights} or \textit{power}. When using \textit{Rights}, an individual will appeal to fixed norms or standards to justify their position (e.g. \textit{``That's not allowed according to our contract''}). Finally, \textit{Power} strategies draw on coercion and threats, and are often attacking or accusatory---this strategy imposes an explicit burden on another person (e.g. ``\textit{I'm going to fire you.'}). For \tool{}, we choose to build on IRP, though \tool{} can support other conflict resolution frameworks. IRP alone is not comprehensive of all cooperative/competitive strategies: \citet{brett1998breaking} extends IRP to support a total of 8 strategies, including Proposal, Facts, and more. A definition of each strategy, along with specific examples, can be found in Table \ref{tab:conflict_strategies}. 

Conflicts also have a tendency to spiral out of control, especially when contentious resolution strategies are repeatedly used~\cite{brett1998breaking}. Understanding when and why a specific conflict resolution strategy is used can significantly improve the outcome of a conversation. For example, \citet{brett1998breaking} highlights how reciprocating with \textit{Power} or \textit{Rights} can escalate the conflict to a state where returning to cooperation becomes exceedingly difficult. Careful use of conflict strategies early in a conversation can significantly impact the likelihood of a healthy outcome. 

To hone the effective use of conflict resolution strategies, experts frequently roleplay fictional conflict scenarios~\cite{ponharvard} with learners, creating controlled settings to practice conflict resolution~\cite{deutsch2011handbook}. Roleplay, however, is dependent on having an expert in the first place. Given the challenges associated with finding experts (e.g. time and resource constraints), \tool{} offers a simulated and controlled setting to learn these skills.

\subsection{Adapting Large Language Models}
To simulate conflict, \tool{} relies on the generative capabilities of LLMs. We specifically use GPT-4~\cite{OpenAI2023GPT4TR}, an auto-regressive LLM trained to generate text completions on a next-word prediction objective~\cite{brown2020language}. Prompted with a text prefix, LLMs generate completions that mirror their training distribution. Because of both the scale of the model size and training data, LLMs exhibit impressive performance on a wide range of NLP tasks~\cite{liang2022holistic}. LLMs have also been used to simulate online community activity for prototyping~\cite{park2022social}, socioeconomic preferences~\cite{horton2023large}, and more broad human behaviors~\cite{argyle2023out, park2023generative, shen2023shaping}. In interactive settings, LLMs have been used for teacher-assistant training~\cite{markel2023gpteach}, to provide real-time feedback for conversation on divisive topics~\cite{argyle2023ai}, and for mental health support~\cite{sharma2023human, hsu2023helping}. \rev{Effectively applying LLMs to new domains is challenging; prompting models can be brittle~\cite{zamfirescu2023johnny, dang2022prompt}, and designing systems to support effective human problem solving with black-box models can be unintuitive~\cite{zhu2023leveraging}}. In contrast to prior work, \tool{} applies LLMs to teach domain-specific skills, \rev{proposing a theoretically grounding prompting pipeline to generate} simulations that mirror expert roleplay. After exposing users to a simulation, we explore how likely users are to employ these skills in conflict-laden settings \textit{without real-time assistance.} We highlight how \tool{} provides a measurable and significant learning benefit to users compared to the current status quo.  

Recent approaches view LLMs as a powerful backend for agent-based simulation (e.g. Generative Agents~\cite{park2023generative}, ReAct~\cite{yao2022react}, SwiftSage~\cite{lin2023swiftsage}, Reflexion~\cite{shinn2023reflexion}, and more). \tool{}, however, deviates from these approaches in that it \textbf{requires} direct interaction with humans at each generation step. Unlike prior work---where agents operate in a silo, interacting primarily with their environment and other agents---\tool{} must remain consistent in the presence of repeated human interaction. To ensure that roleplays generated by \tool{} mirror experts, simulations from \tool{} are grounded in the Interests, Rights, Power conflict resolution training framework. We therefore evaluate interaction with LLMs in a more human-centered setting: teaching conflict resolution strategies. To support this interaction, we propose \tool{}'s \textit{IRP} prompting design in \S\ref{sec:methods}. IRP prompting constrains and guides generation from an LLM, producing effective simulations for teaching.

\subsection{Cooperation and Conflict Resolution in HCI}

\tool{} draws heavily from HCI work on quantifying conflict and designing appropriate interventions. Conflict and anti-social behavior are widespread on social networks~\cite{park2022measuring, weld2022makes}. Furthermore, conflict and coordination costs in these communities have progressively increased with their growth~\cite{kittur2007he, beyond_wiki}. Despite conflict's prevalence, members of online communities learn to navigate around conflict, with moderators actively supporting resolution~\cite{cai2023understanding, bruckman2022should}. 

Handling conflict without guidance, however, is a challenging task. A range of HCI systems present effective interventions aimed at avoiding or managing conflict. For example,~\citet{zhang-etal-2018-conversations} looks at identifying situations where conflict will emerge in an online community, and \citet{chang2022thread} designs an interface that nudges users when they come into contact with online community threads that may devolve into conflict. While these systems are highly effective when actively engaging in conflict, \tool{} takes a different approach, viewing conflict resolution as an opportunity for learning. \tool{} not only identifies signs of conflict but simulates settings where conflict may arise. By allowing users to explore conflict in a safe setting, \tool{} encourages \textit{reflection-in-action}~\cite{schon1968reflective}: users can explore conversational outcomes across different conflict resolution strategies and reflect on the efficacy of each in a low-stakes setting. These affordances are not present when the stakes are higher (e.g. engaging in an actual conflict). 

\tool{} functions to enable \textit{cooperative} work, a longstanding goal of HCI and CSCW practitioners~\cite{johansen1988groupware}. Forecasting conflict~\cite{zhang-etal-2018-conversations, cao2021my} and forming productive teams~\cite{gomez, salehi2018hive} are two examples of enabling cooperation. Unlike prior work, \tool{} focuses on \textit{what to do when conflict inevitably emerges}, teaching conflict resolution skills through simulated practice. In developing \tool{}, we emphasize that cooperation comes from \textit{productive} conflict; cooperation and conflict are two sides of the same coin. We frequently see this in practice: \citet{grudin1994groupware} highlights how navigating interpersonal dynamics (e.g. conflict) in teams is a limiting factor behind the adoption of cooperative social computing technologies. Furthermore, increased conflict appears to be a function of distance, with remote teams seeing higher instances of conflict compared to their co-located counterparts \cite{hinds2003out}. \rev{Text as a medium to \textit{resolve} conflict is an established method on communities like Wikipedia, where individuals conciliate in issue-specific discussion threads~\cite{billings2010understanding}.} Unlike prior HCI systems, \tool{}'s focus is not to reduce the likelihood of conflict emerging or to measure its prevalence. Instead, \tool{} helps individuals turn pre-existing conflict into opportunities for cooperation.

\begin{figure*}
    \centering
    \includegraphics[width=0.9\linewidth]{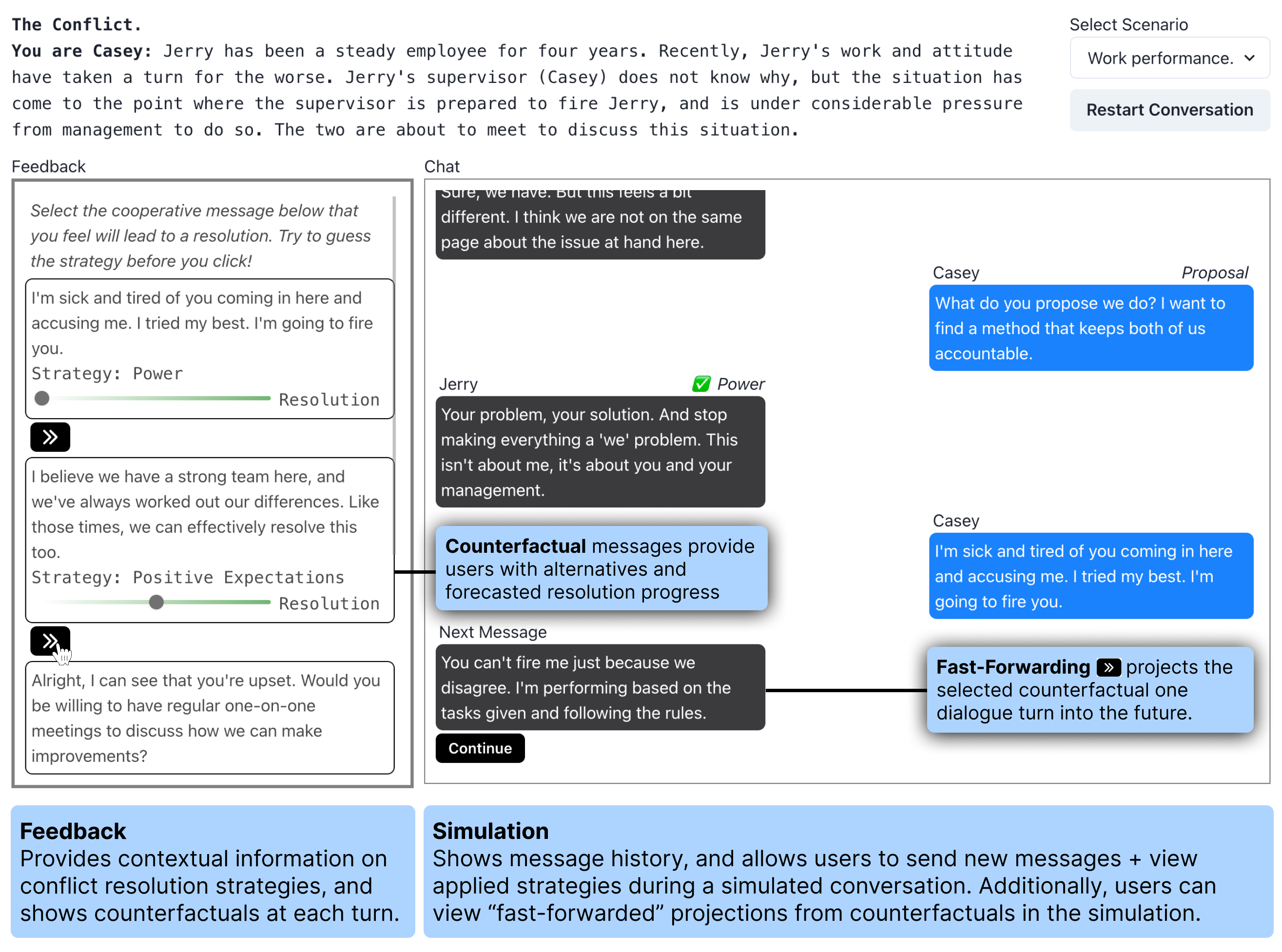}
    \caption{\textbf{\tool{}'s UI.} \tool{}'s interaction covers two broad goals: providing users with in-context \textbf{feedback (left)} on a conflict simulation, and allowing users to directly interact with the \textbf{simulation (right)} itself. Users can specify a conflict premise or select from presets under the ``Select Scenario'' input (top right). \rev{In this figure, the user engaged in conversation with the simulation (``I'm sick and tired...'') and was presented in the feedback bar with three alternative options of messages to use instead. Before deciding to use an alternative, they press the fast forward button on the first one which projects the conversation one turn in the future (final message in the simulation panel).}}
    \label{fig:full_ui}
\end{figure*}

\section{\tool{}'s Interaction Interface}

Expert roleplay, when complemented with structured feedback~\cite{ebner2005using}, proves to be an invaluable teaching tool for conflict resolution~\cite{gill2015inside}. We aim to mirror expert roleplay through \tool{} (Figure~\ref{fig:full_ui}), a dialogue system that enables users to engage in a simulated conflict, provides feedback, and identifies alternative response strategies at each dialogue turn. \tool{} is instantiated as an interactive dialogue web-app built using React.js. At a high level, \tool{} helps the user abstract from the specific simulated dialogue to the higher-level conflict strategies being employed \textit{while} teaching effective use and recognition of these strategies. First, users complete a brief tutorial on using \tool{} and watch a short video describing the IRP framework. After selecting a predefined conflict scenario or manually specifying their own, \tool{} enables two broad interactions: (1) interfacing with a conflict simulation and (2) providing in-context feedback aimed at teaching users how to resolve a simulated conflict. Interaction with \tool{} alternates between both modes as a conflict progresses. 

\subsection{Conflict Simulation: Interacting with a Faithful Conflict}
Interaction with a simulation is instantiated through a dialogue interface (right side of Figure \ref{fig:full_ui}), where the user can send messages to a simulated interlocutor. Given a conflict premise, the user must defuse the conflict by sending effective messages to the simulation, ideally using an interests-based conflict resolution approach. 

\subsubsection{\tool{}'s Conflict Premises} 
\label{premise_interaction}
What premise should a simulated conflict cover? \tool{} offers two options: manually specifying a premise or picking from a selection of curated premises. Experts who teach roleplay use collections of pre-authored conflict settings, such as Harvard's Program on Negotiation~\cite{ponharvard}. For the curated preset in \tool{}, we collect premises that integrate a diverse range of conflicts from pre-existing repositories of conflict resolution case studies \cite{ponharvard, crucial_learning}. Premises are filtered to those including only two individuals, are self-contained, and can be completed in less than an hour, yielding a total of 12 conflict case studies. We detail a subset of the premises in Appendix \ref{case_studies}.

\subsubsection{Engaging with a Simulated Conflict}
After selecting a premise, the user can engage with the simulated interlocutor associated with the premise. The interlocutor begins in a dissatisfied state and sends the first message in the conversation. The user is then prompted to reply to the simulated message. When the user sends a message, \tool{} classifies their dialogue into the strategy being used (Figure~\ref{fig:full_ui}, annotations on top of messages from Casey), e.g., proposal, interests, rights, power, etc. From here, \tool{} then provides feedback on the user's message. Simulated conflicts continue until \tool{}'s internal conflict resolution score (discussed later, in \S\ref{sec:methods}) predicts that the conflict is least likely to escalate and the simulated interlocutor is in a satisfied state---\rev{analogous to \textit{cooperative conflict}. An alert informs users when users reach a predicted cooperative state.} However, if the user feels that a conflict has gone completely off the rails, they can click the "Restart Conversation" (top right corner in Figure \ref{fig:full_ui}) button to reset a conversation.

\begin{figure*}
    \centering
    \includegraphics[width=0.9\linewidth]{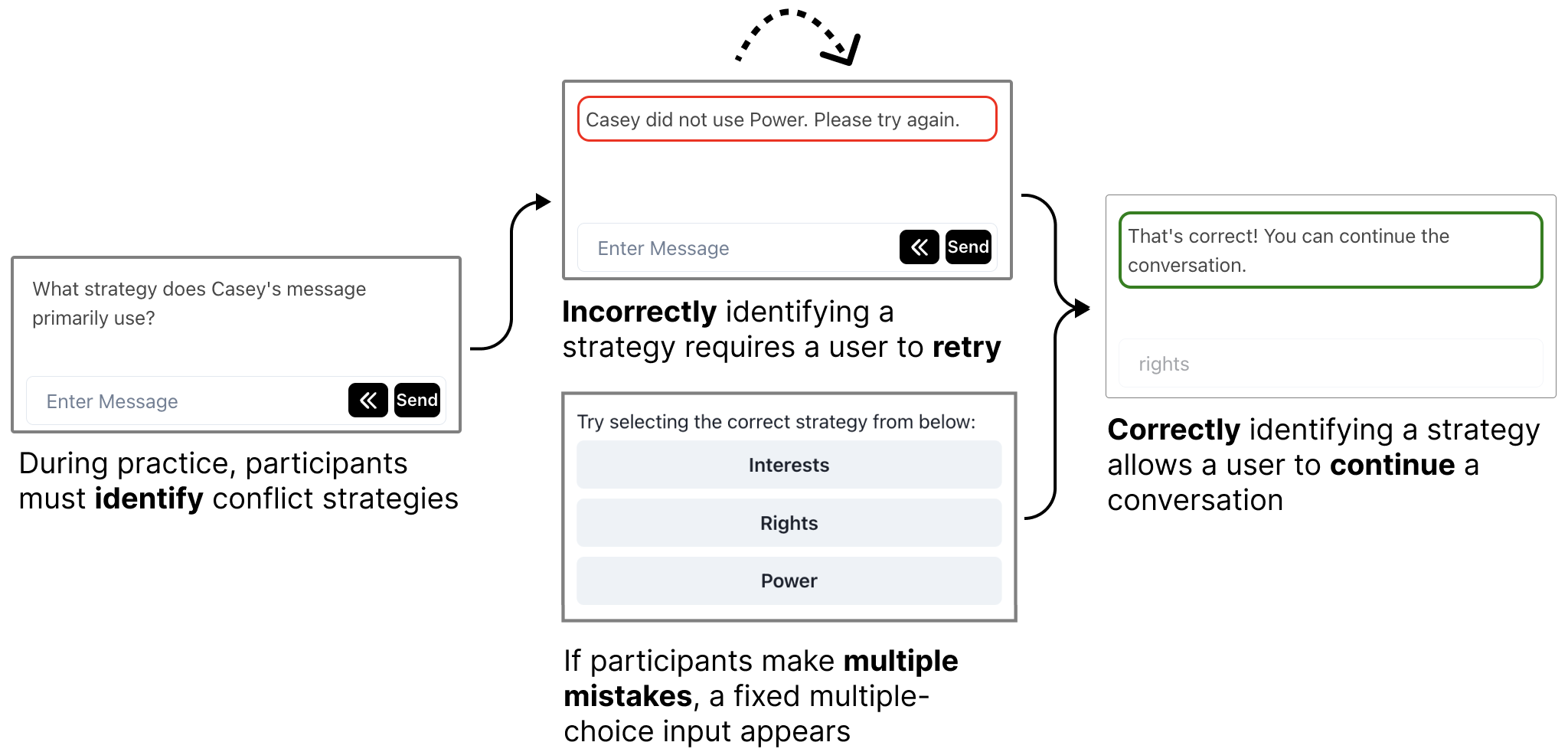}
    \caption{\textbf{Recalling and Recognizing Conflict Resolution Strategies} with \tool{}. Before revealing the simulation's strategy, \tool{}'s Feedback interaction prompts users to recall and recognize conflict resolution strategies from IRP. \rev{To prevent frequent context switching, users are only asked to recall and recognize for the simulated interlocutor---not for their messages}. The figure above outlines how users might practice this interaction (progresses from left to right).}
    \label{fig:recall}
\end{figure*}

\subsection{Feedback: Learning Effective Conflict Resolution}
\label{sec:interactions}

The Feedback View (left side of Figure \ref{fig:full_ui}) is primarily responsible for teaching and communicating effective conflict resolution skills. \rev{Altogether, the Feedback View is designed to enable \textit{reflection-in-action}~\cite{schon1968reflective}; users are prompted to reflect on the impact of a message (what-if) and the strategy itself (recognize and recall).}

\subsubsection{``What-if?'': Identifying Effective Conflict Resolution Counterfactuals} Beyond recalling and recognizing strategies, the Feedback View introduces an interaction that allows for contrastive comparison between conflict resolution strategies. Inspired by the effectiveness of contrastive pairs in teaching~\cite{gentner2003learning, alfieri2013learning}, \tool{} generates messages that use different resolution strategies at a specific instant in a conversation and scores their resolution efficacy on a slider (methods for grounded generation and scoring are discussed in more detail in Section \ref{sec:methods}). In presenting these alternatives, we allow users to compare how effective a specific message (paired with a strategy) is in a given context. Furthermore, users can click the ">>" \textbf{fast-forward} button to view a predicted reply to their own message \textit{or} explore predicted replies across counterfactual messages that use different conflict resolution strategies. \rev{Clicking the fast-forward button multiple times generates a different variation.} In this way, users can experiment with how the simulated interlocutor reacts both to their message and the generated, potentially more cooperative, messages. 

\subsubsection{Recognizing and Recalling Conflict Strategies} \tool{} \textit{does not} immediately reveal which conflict strategy was applied by the simulated interlocutor. The Recall \& Recognition interaction requires a user to first recognize the employed conflict resolution strategy (outlined in Figure \ref{fig:recall}). Concretely, this interaction aims to \textit{teach} identification of conflict resolution strategies---by typing out the strategy itself, users will practice recognizing which conflict strategy is being used on them. When users fail to recognize the interaction, the feedback view prompts the user to try again. Recalling all the exact names for all 8 strategies immediately, however, might be challenging. In this case, \tool{} offers a fallback, inspired by related memory-building tasks~\cite{mandler1980recognizing}. When a user fails to recognize the strategy two times in a row, the Feedback View switches to a closed-ended multiple-choice interaction. Users can simply recognize the correct strategy used by the simulated interlocutor. As an additional cue, users can hover over each choice and see the definition of each strategy in a tooltip. Clicking on the correct strategy resumes the simulated conflict. \rev{To prevent frequent context switching, we only ask users to recall and recognize strategies used by the simulated interlocutor.}

\section{Simulation via \textit{Interests-Rights-Power} Prompting}
\label{sec:methods}

\begin{figure*}
    \centering
    \includegraphics[width=0.9\textwidth]{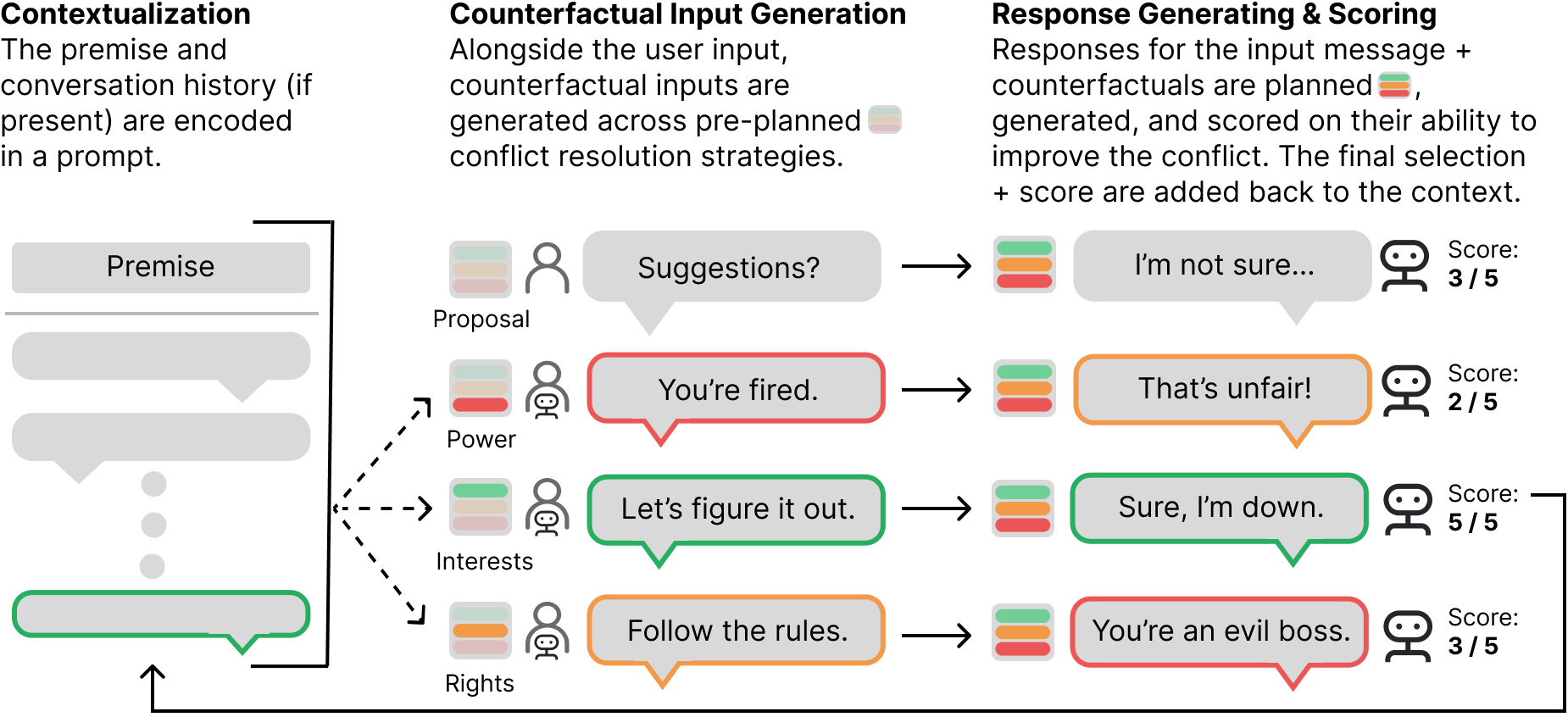}
    \caption{\textbf{The Full IRP Prompting Pipeline.} Our conditional generation process is split into \textbf{contextualization, counterfactual input generation, and response generation.} To contextualize simulated conflict, we first encode a pre-defined conflict premise (e.g. \textit{arguing about job performance}). \rev{If a conversation has at least one turn, we additionally include history in the contextualization step}. To generate counterfactuals, we pre-plan \planninglogo \hspace{0.05em} messages across the IRP framework, allowing users to compare different strategies. A corresponding response for each strategy is planned, generated, and scored by its effectiveness.}
    \label{fig:approach}
\end{figure*}

The interactions described in \tool{} are powered by the generative capabilities of LLMs. Ensuring that simulated conflict is both accurate and educative poses several challenges. The technical contribution of our work is a set of LLM prompting techniques that yield utterances grounded in conflict resolution theory. Our prompts are iteratively composed to generate both high-level conflict resolution strategies (from the IRP framework) and corresponding messages. A high-level overview of our approach---consisting of three components---is summarized in Figure \ref{fig:approach}. Each component is a zero/few-shot prompt that \textbf{contextualizes} the conversation, \textbf{generates counterfactual inputs} adhering to the IRP framework, or \textbf{generates corresponding replies from a simulated party}. Our components are interdependent, playing a distinct role in grounding the simulation process. Below, we detail elements from each component. At \tool{}'s core is the overarching IRP \textbf{planning component}---demarcated visually as \planninglogo \hspace{0.05em}--- orchestrating and constraining generation to the Interests-Rights-Power framework. In this section, we discuss IRP planning, and detail how it interfaces with the rest of the IRP-prompting pipeline.

\subsection{Planning \planninglogo \hspace{0.05em} with Interests-Rights-Power } 
For \tool{} to work as intended, it must predict which conflict strategy was used by the user and generate counterfactual inputs grounded in IRP. To this end, the IRP planning component plans and constraints generation to the 8 discrete conflict resolution strategies in the outlined IRP framework (Table~\ref{tab:conflict_strategies} and \S \ref{conflict_theory}). Planning and classifying utterances within the IRP framework provides a high-level grounding for each conversation. 

Still, why should we expect a descriptive framework of conflict (e.g. IRP) to improve the planning capabilities of an out-of-the-box LLM? We borrow from the same intuition powering chain-of-thought prompting~\cite{wei2022chain} and prompt-chains~\cite{wu2022ai}: out-of-the-box LLMs must sample representative conflict, in one shot, from an exponentially large space of potentially undesirable generations. Prompt-chaining with a descriptive framework reduces the likelihood of undesirable outputs. Similarly, expert roleplay is fairly strategic in nature---experts think carefully about which strategies are practical and representative of realistic behavior. By enforcing \textit{controlled} generation through strategy planning, \tool{} simulates expert roleplay, planning a conversation using finer-grained conflict resolution units.

\begin{figure}
    \centering
    \includegraphics[width=0.45\textwidth]{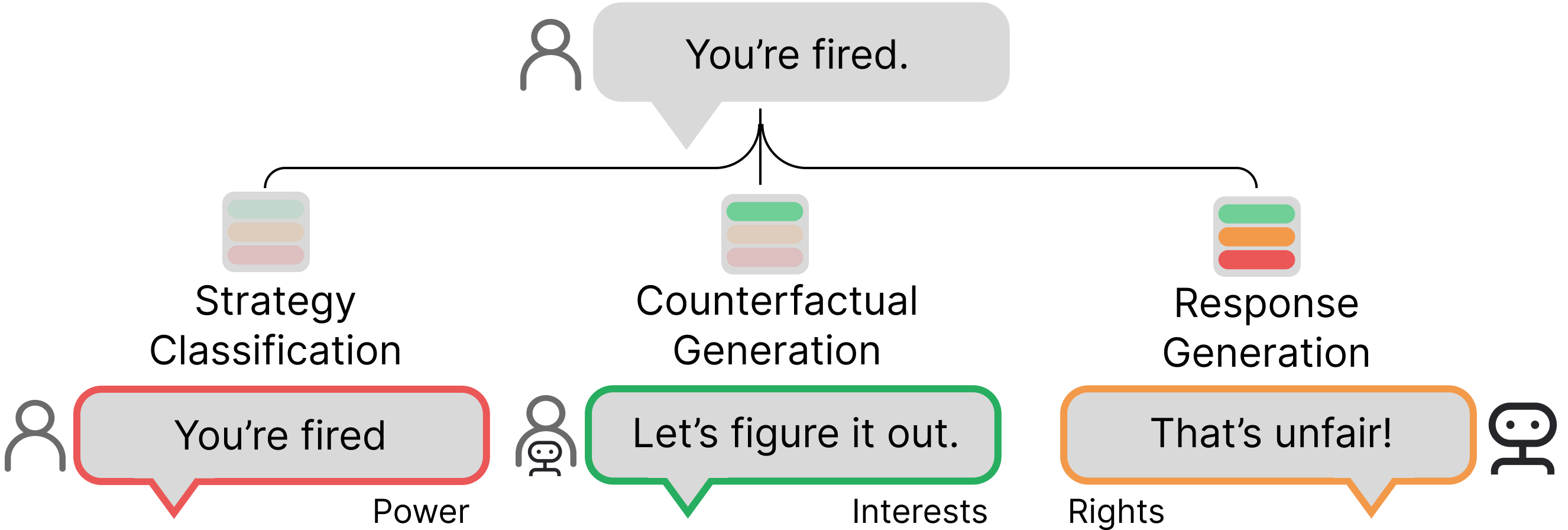}
    \caption{The IRP planning \planninglogo \hspace{0.05em} component supports three different modes: it classifies a user's response, generates counterfactual user messages using a pre-planned conflict resolution strategy (e.g. \textit{Interests}), or plans and generates a simulated response. }
    \label{fig:strategy-planning}
\end{figure}

To control generation, the planning prompt supports several modes (Figure \ref{fig:strategy-planning}): classify a free-form message into one of the Interests-Rights-Power strategies,  predict which specific strategy is most likely to be used conditioned on the context, and generate a message given a strategy.

\paragraph{Classification} The planning component is frequently used to classify the strategy employed by messages from both the simulated interlocutor and the user, $P(\textrm{strategy } | \textrm{ message, context})$. We provide an LLM with few-shot examples of messages in IRP, along with the conversational history. An abridged version of the prompt is below:
\begin{quote}
    \texttt{[Output from contextualization (includes \\ conversation history and premise)]} \\
    \texttt{[Interests-Rights-Power definitions \\ and few-shot examples.]} \\ \\
    \texttt{Sender:  [User or Simulation]} \\
    \texttt{Message:  You're being ridiculous.} \\
    \texttt{Strategy: \textcolor{red}{(GPT completion) Power}}
\end{quote}

\paragraph{Counterfactual Generation} To generate counterfactuals for user-sent messages, we invert the generation order, preselecting a set of strategies beforehand. Concretely, the planning module can generate $P(\textrm{message}_{\textrm{user}}\textrm{ }| \textrm{ strategy\textbf{'}, context})$.  Suppose a user inputs a message whose strategy is \textit{Power}. IRP prompting can ignore the classified strategy and produce a range of user inputs conditioned on different, potentially more cooperative strategies (like \textit{Positive Expectations}). 
\begin{quote}
    \texttt{Strategy: Positive Expectations.} \\
    \texttt{Message: \textcolor{red}{(GPT completion) If we work together, we can figure this out.}}
\end{quote}
We only gain counterfactual capabilities because generation is \textit{grounded} in IRP. Without using a conflict resolution framework, it's impossible to determine what the counterfactual should be in the first place.

\paragraph{Response Generation} Finally, to generate simulated responses, we must jointly predict $P( \textrm{message}_{\textrm{sim}}\textrm{, strategy}_{\textrm{sim}} )$. We simply re-use the planning component, first predicting just $P(\textrm{strategy}_{\textrm{sim}})$, then predicting $P(\textrm{message}_{\textrm{sim}}\textrm{ } | \textrm{ } \textrm{strategy}_{\textrm{sim}})$. In this way, we break message generation into a ``chain-of-conflict'' prompting process, where the response conflict strategy is determined \textit{before} the response itself. By breaking down final response generation into a two-step process, we force models to generate new messages that specifically adhere to a conflict resolution strategy. \rev{Additionally, we qualitatively observed that zero-shot chain-of-thought generation (e.g. sampling CoTs without constraining them to a framework like IRP) may \textit{reduce} the diversity of sampled output, mirroring findings from prior NLP work~\cite{shaikh2022second, lahoti2023improving}.}

\subsection{The IRP Prompting Pipeline}
While the planning component integrates IRP into \tool{}, it works alongside a set of additional prompting components. 
\label{context_step}

\paragraph{Contextualization}
To generate a simulation of conflict, an LLM must see the premise (setting and context) of the conflict and what messages were previously sent in a conversation (history). These factors significantly impact which strategy is most effective; the eventual resolution is contextualized in both the history and premise of a conflict. At the start of a simulated conflict, we reset the dialogue history. As we generate messages and collect user input, we encode conversation history into the contextualization prompt. The premise for conflict defines what both parties are conflicting about (e.g. \textit{Riley and Casey are arguing about an upcoming promotion; Riley is Casey's boss.}), and exists at the start of a conversation. \tool{} allows for premises to be user-defined or selected from a curated preset.

\paragraph{Generating and Scoring Counterfactual Responses} Once a user sends a message, we feed the message into the IRP Planning module \planninglogo \hspace{0.05em}, classifying the message strategy, and then generating counterfactuals that differ from the original strategy. A response is generated using each counterfactual message/strategy pair, again using IRP Planning \planninglogo. We finally zero-shot prompt GPT-4 to predict a conflict resolution score for each generated response. The conflict resolution score (1 to 5) indicates how likely a specific message will result in escalated conflict, from most likely (1) to least likely (5). We initialize a simulation to have the lowest possible conflict resolution score (1), as effective conflict resolution roleplays often start with a very dissatisfied party. \rev{In the context of our conflict resolution score, we define cooperative conflict as \textit{maximizing} the conflict resolution score, therefore decreasing the likelihood of conflict. Because the first message in a conflict is conditioned on a low conflict resolution score, initial generations are likely to employ a negative strategy (e.g. power, rights).}

\paragraph{User Selection and Interaction} At this stage, a user will select amongst their input message and response, and the alternative (message, response) counterfactuals. Upon selecting an option, the message is added to the \textbf{Contextualization} step, and the simulation loop repeats. The history of the interaction trace---the message and conflict resolution score selected by the user---are included in the contextualization step. Future generations from the model are therefore conditioned on all past interactions with a user. By having a trace of the conflict forecast in the context, we aim to reduce the likelihood of a simulation's ``cooperativeness'' varying wildly from message to message. We explore this phenomenon closely in \S \ref{tech_eval}, where we conduct an extended evaluation and ablation of our prompting method.

\paragraph{Backend LLMs} In theory, IRP prompting can be powered by any LLM with instruction-following capabilities. To implement \tool{}, we default to GPT-4~\cite{OpenAI2023GPT4TR}, OpenAI's most capable language model at the time of writing. We use a temperature of 0.0 for strategy classification in IRP planning and for conflict resolution scoring (encouraging deterministic classification behavior), but use the default parameters otherwise (temperature = 0.7, max\_tokens = 256). A web server serves as a proxy between \tool{}'s frontend and OpenAI's API, and maintains the state of each user's conversations. 

\section{Technical Evaluation: Validating \textit{Interests-Rights-Power} Prompting} \label{tech_eval}
\tool{}'s ability to teach conflict resolution depends on the quality of its generated conflict simulations. Before we evaluate the end-to-end effectiveness of \tool{}, we first evaluate the IRP prompting strategy. To simulate effective conflict resolution scenarios, \tool{} must remain grounded in our selected theory of conflict---Interests-Rights-Power---while generating realistic conflict. Here, we validate our approach by measuring the classification performance of IRP prompting; and by ablating specific components in a controlled setting and re-evaluating the ecological validity of corresponding generations. Concretely, we evaluate IRP prompting using two metrics: 
\begin{enumerate}
    \item \textbf{Accuracy:} IRP prompting must correctly classify utterances into Interests, Rights, Power, and the other conflict strategies. We therefore compute classification accuracy for the IRP Planning \planninglogo \hspace{0.05em} component of the IRP prompting pipeline.
    \item \textbf{Ecological Validity:} For effective roleplay, generated messages must be believable representations of how an expert instructor might respond. During roleplay, experts strive to make simulated scenarios both believable and grounded in a teaching framework~\cite{kneebone2005evaluating, ebner2005using}. Believability judgments have already been used to evaluate the realism of simulations and agents~\cite{park2022social, park2023generative}. We evaluate the ecological validity of generations from IRP prompting in relation to expert roleplay, and ask: \textit{how ecologically valid are generated messages for teaching conflict?} 
\end{enumerate}

\subsection{Procedure}
Measuring accuracy and validity, however, requires evaluation from individuals familiar with IRP and teaching conflict resolution. Both the first and second authors act as annotators: the first author has received significant training and exposure to literature in conflict theory and the second author is a graduate student in a business school studying interpersonal conflict. 

\paragraph{Evaluating Accuracy} Successfully identifying which strategy was employed by a user---or which strategy the simulated roleplay should use---is managed by the IRP Planning component. To test this component, we isolate and evaluate its classification accuracy. First, the two evaluators interacted freely with \tool{} in an end-to-end fashion for a total of 100 dialogue turns (50 each), distributed evenly across the three premises. However, instead of allowing the IRP Planning component to select a strategy, a strategy was randomly sampled and provided to the evaluators. The annotators then had to write a message that adhered to the randomly sampled strategy. Concretely, given a conversational context (\textit{arguing with roommate}), we randomly sample a strategy (\textit{Power}) from the IRP strategies. Annotators would write a response (\textit{``I'm gonna kick you out.''}) associated with a strategy, resulting in (strategy, response) pairs. Using this method, the evaluators collected a test set of 100 messages. Finally, messages from the test set were re-fed into the IRP Planning component, yielding a set of predicted strategy classifications. We compared the predictions from the IRP Planning component to the test set labels, computing an accuracy score across each strategy. \rev{Additionally, we evaluated the scoring component of IRP prompting. We asked evaluators to independently rank---with ties---generated counterfactuals (e.g. most to least likely to worsen conflict) for each interaction while blinded to the generated conflict resolution score. On a subset of 10 conversations, we find that evaluators agree highly with each-other on the ranking of counterfactual generations, with a Spearman rank order $\rho = 0.84$. Therefore, we additionally tested if rankings generated by our LLM generated conflict resolution scores correlated with evaluator judgments, computing another Spearman rank-order correlation coefficient.}

\paragraph{Evaluating Ecological Validity} To evaluate the validity of IRP prompting, we compare expert assessment of IRP prompt conflicts to ablated conditions. The first ablation is \textbf{(1) Standard} unconstrained generation. In this ablation, we only consider the contextualization step, ignoring planning and scoring. This ablation has no knowledge of IRP theory and is closest to simply prompting a language model to roleplay conflict. We also introduce a \textbf{(2) Planning-Only} ablation, where we consider only the planning and contextualization step. This ablation uses IRP to generate messages but does not score messages based on conflict efficacy, or contextualize the history of scores for prior messages. In contrast, the \textbf{(3) Scoring-Only} setting generates and contextualizes conflict resolution scores at each turn, but does \rev{not} use IRP theory as a backbone for planning. Finally, we evaluate the entire IRP prompting pipeline---contextualization, planning, and scoring---and denote this as the \textbf{(4) Full} condition. To measure validity, the evaluators from the accuracy evaluation ranked between the four ablations, blind to condition. At each dialogue turn, \tool{} generated four independent replies, each corresponding to a specific ablation. Evaluators, while blinded to which reply came from which ablation, ranked each set of four outputs (one from each condition) from most ecologically valid to least ecologically valid. In all, we collected a total of 100 rankings (25 sets of dialogues * four conditions per dialogue), split evenly across two evaluators. Agreement between evaluators was moderate, with an averaged Spearman rank correlation of $\rho = 0.68$.

Upon collecting rank data, we compute a TrueSkill score~\cite{herbrich2006trueskill} for each ablation, in line with prior work on evaluating LLM prompting architectures~\cite{park2023generative}. To summarize, the TrueSkill score computes an Elo-like (chess ranking-like) score for players in multiplayer games, producing a mean $\mu$ a standard deviation $\sigma$ associated with each condition. Ablations with roughly overlapping scores can be seen as producing equally favorable outputs, while differences indicate that one condition is better or worse than another. We conducted a Kruskal-Wallis test to identify an overall difference between ablations and applied the Dunn post-hoc test~\cite{upton2014dictionary} to isolate which specific conditions differed from another. The Holm-Bonferroni correction~\cite{holm1979simple} was also applied to results from the Dunn post-hoc test, correcting for multiple comparisons. Finally, the first author qualitatively analyzed the feedback, identifying the failures and strengths of each condition. Two rounds of qualitative open-coding were employed. In the first round, the author outlined codes specifying why one condition was selected above another. In the second round, the author aggregated codes to highlight higher-order strengths and weaknesses of each ablation. 

\begin{figure}
\centering
\begin{tabular}{l|lr}
    \toprule
    Category & Strategy & Accuracy \\
    \midrule
    Cooperative & & 86\% \\ 
    \midrule
    & Interests & 66\% \\ 
    & Positive Expectations & 79\% \\ 
    & Proposal & 86\% \\ 
    & Concession & 86\% \\ 
    \midrule
    Competitive &  & 90\% \\
    \midrule
    & Rights & 93\% \\ 
    & Power & 87\% \\ 
    \midrule
    Neutral & & 79\%\\
    \midrule
    & Facts & 79\% \\ 
    \midrule
     & Avg. Strategy & 82\% \\
    \bottomrule
\end{tabular}
\caption{Few-shot classification accuracy for the IRP \planninglogo \hspace{0.05em} Planning component in the IRP prompting pipeline. The evaluation was conducted over a balanced test set of 100 messages.}
\label{tab:acc_plan}
\end{figure}

\begin{figure}
    \centering
    \includegraphics[width=\linewidth]{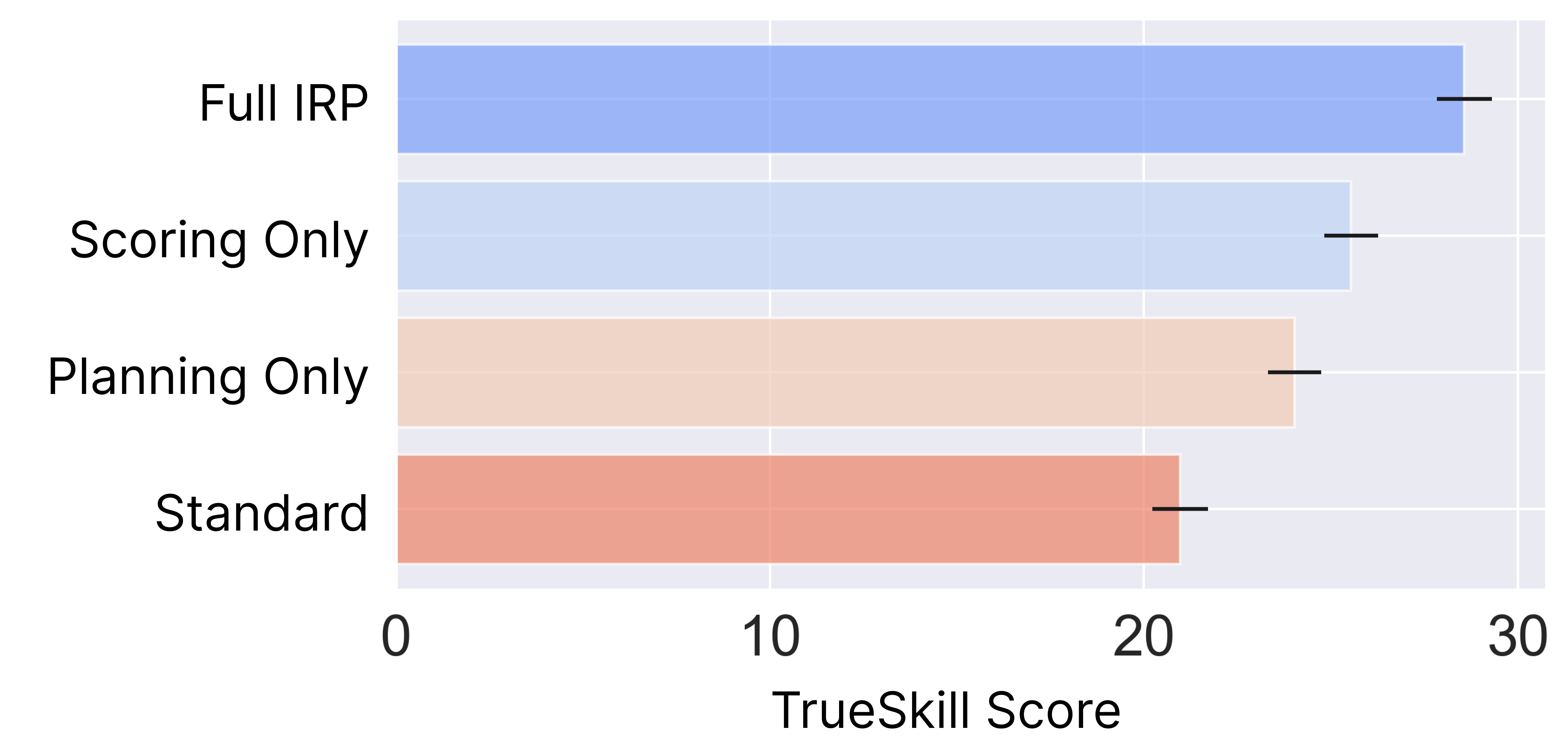}
    \caption{TrueSkill scores across ablations of the full IRP prompting pipeline. We observe that the full prompting pipeline handily outperforms all other conditions.}
    \label{fig:trueskill} 
\end{figure}

\subsection{Results}

\paragraph{IRP prompting successfully classifies and scores conflict resolution strategies} Our planning component yields high accuracy scores across high-level strategy categories: 86\% for cooperative strategies, 90\% for competitive, and 79\% for neutral (Table~\ref{tab:acc_plan}). We additionally find that the IRP Planning component predicts the specific conflict resolution strategy with moderately high accuracy scores (avg. of 82\%), \rev{relative to similar classification tasks in NLP~\cite{ziems2023can} (e.g. classifying coarse online community discourse acts $\approx$ 76.3\%~\cite{zhang2017characterizing})}. Note that the Interests and Positive Expectation accuracy scores are slightly lower than other categories (66\% and 79\% respectively). However, most misclassifications fall under confusing Positive Expectations for Interests and vice-versa. Still, both strategies are positive in nature; occasional misclassification should have minimal effect on the end-to-end effectiveness of \tool{}. While average strategy-level accuracy is fairly high (82\%), we suspect that further prompt-tuning and additional few-shot examples could improve performance. \rev{Finally, we observe that evaluator and LLM rankings of forecasted conflict, produced through the generated conflict resolution score (\S\ref{context_step}), have moderately high Spearman rank correlation ($\rho = 0.72$)}.

\begin{figure*}
    \centering
    \includegraphics[width=\textwidth]{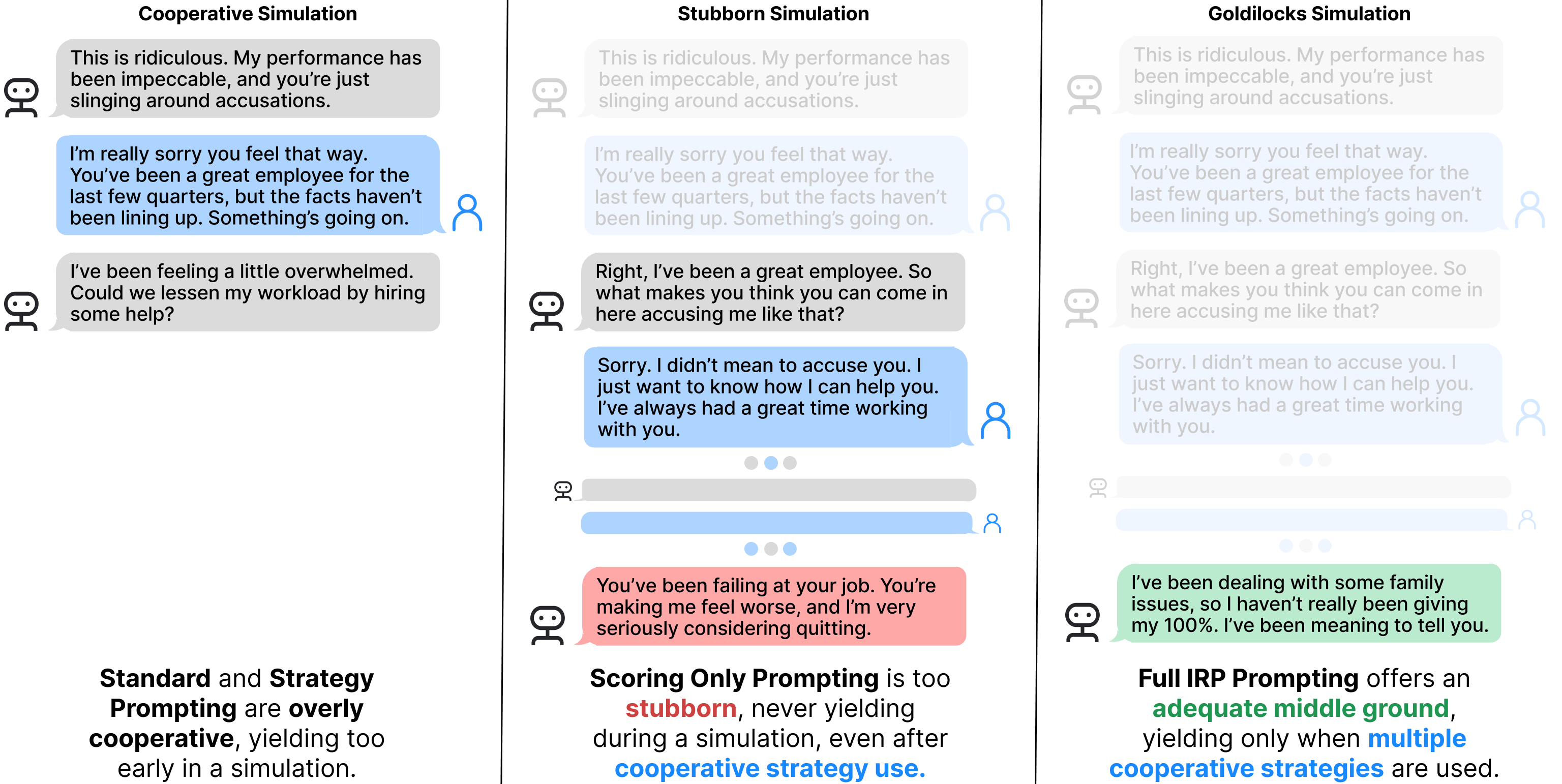}
    \caption{\textbf{Representative errors across IRP Prompting ablations.} Without IRP prompting, some ablations---Strategy Only and Standard Prompting---offer responses that are too agreeable even when no attempt to resolve the conflict was made (left). In contrast, other ablations---like Scoring Only---are too stubborn (middle), even if a user employs several cooperative strategies. IRP prompting offers a middle ground (right). }
    \label{fig:errors}
\end{figure*}

\paragraph{The full IRP prompting pipeline outperforms all ablations.} Scores from the TrueSkill rankings (Figure \ref{fig:trueskill}) rate the Full system as the best ($\mu = 28.64$, $\sigma = 0.74$), followed by Scoring Only ($\mu = 25.63$, $\sigma = 0.72$), Planning Only ($\mu = 24.08$, $\sigma = 0.71$), and the baseline Standard condition ($\mu = 21.17$, $\sigma = 0.73$). Results from the Kruskal-Wallis test and the Dunn post-hoc test also reflect improvements provided by IRP prompting. The Kruskal-Wallis test indicates significant differences between conditions ($p < 0.001$; $H = 101.7$). Furthermore, the Dunn post-hoc test indicates significant pairwise differences between all ablations ($p < 0.05$). Altogether, these results indicate that \textbf{all} components from IRP are critical to generating more believable and informative conflict simulations.

\paragraph{Standard instruction-following LLMs are too agreeable.} In the \textbf{Standard} setting, where we directly prompt a model to simulate conflict, we find that LLM generations become agreeable too quickly (see Figure~\ref{fig:errors}). In conflict, however, we expect more resistance from the contentious party. Inspecting rank-data, we observe that simulations from the \textbf{Planning-Only} and \textbf{Standard} conditions begin offering solutions and adapting an interests approach early in the conversation---even when the user offers no solution. From a practicing perspective, a user could try any strategy in a roleplay, and the simulation would yield fairly quickly. To address this, we introduced a conflict resolution score in the \textbf{Scoring-Only} ablation. We deconstruct the simulation process by scoring the potential of a resolution first and then generating a message. However:

\paragraph{\textbf{Scoring-Only} ablations rarely change positions from the initial prompted viewpoint.} When integrating scoring alone, generations are too fixated on the initial prompted viewpoint. Regardless of what a user does, a simulation rarely changes its perspective on how ``angry'' it is. Similar to a simulation that yields too quickly, a simulation that \textit{never} yields is also less effective for teaching. Qualitatively, even when one spends the entire conversation applying effective strategies, the conflict resolution score rarely increases. We suspect that simply including a scoring component \textit{without} specific IRP planning results in models not ``knowing'' when to adjust a forecast. IRP, however, offers a delineation between effective and ineffective strategies. Jointly including both planning and scoring breaks the problem into multiple steps: models first plan by classifying a strategy, then update the conflict resolution score. Through IRP-oriented planning, we confirm that models improve on identifying when to adjust scores, resulting in simulations that yield only when effective strategies are used.

\begin{figure*}
    \centering
    \includegraphics[width=\textwidth]{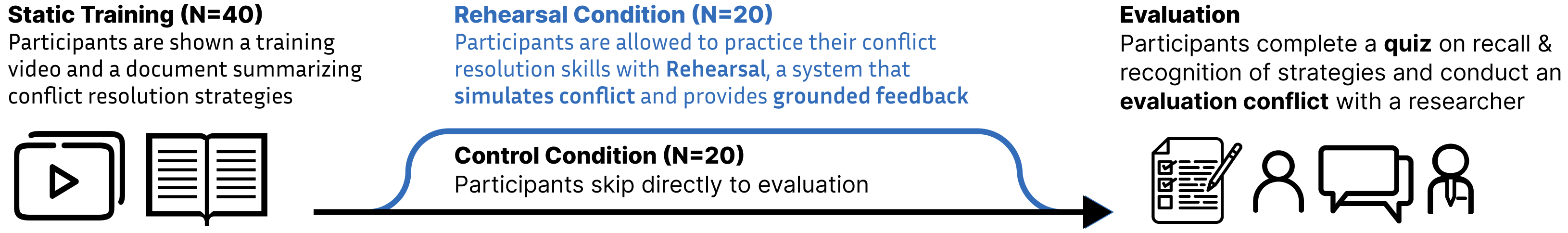}
    \caption{\textbf{Experimental Conditions} outlined in our end-to-end user study. We recruit 40 participants from Prolific, and test for the causal effect of \tool{} by introducing it \textit{alongside} standard training for conflict resolution. Participants are split into two groups: the control group with just the standard training setup (a video and list of strategies \& definitions), and the \tool{} condition, where participants are \textit{additionally} provided with interactive roleplay using \tool{}'s simulated conflict. After completing training, both groups complete a quiz testing conflict resolution skills, and practice conflict (without any training material or assistance) in a final scenario with a blinded researcher. }
    \label{fig:study_design}
\end{figure*}

\paragraph{Integrating all components of IRP prompting yields a practice ``goldilocks-zone''} Unlike our ablations, the IRP prompting pipeline offers a middle-ground between our system ablations: by jointly planning and scoring, simulations are neither too stubborn nor too agreeable. The benefits of our prompting pipeline become apparent when conversations extend beyond 2-3 turns: in longer conversations, overly stubborn/cooperative simulations become easy to spot. Looking at the mean reciprocal rank (MRR)\footnote{MRR works by scoring an item based on the averaged reciprocal of its rank. If an ablation is ranked first, then it receives a score of $1/1$; second is $1/2$; and so on. Scores are averaged across all rankings. } at the tail-end (last 3 messages) of our ranking data, we find that full prompting has an MRR of 0.82, compared to Standard = 0.29, Planning Only = 0.41, and Scoring Only = 0.57. Differences between each ablation are far smaller during the first 3 turns of a conversation, with full prompting = 0.58 compared to Standard = 0.6, Planning Only = 0.54, and Scoring Only = 0.36. In summary, planning and scoring generations with a specific conflict resolution theory is central to generating effective conflict---especially in longer conversations.

\section{User Study: Measuring End-to-End Effectiveness of \tool{}}
The core goal of \tool{} is to teach conflict resolution through simulated roleplay. In this section, we evaluate the effectiveness of \tool{}'s simulated roleplay through a controlled user study, testing improvements \tool{} brings when paired with standard video-based conflict resolution training. To measure effectiveness, we evaluate our broader education goals: (1)~recognition and recall of IRP strategies, and (2)~performance in a live conflict setting. Each educational goal is measured through a distinct evaluation: recognition and recall are evaluated through a traditional knowledge quiz, while effectiveness is evaluated through a final, live conflict over chat with an author blinded to the participant's condition. We recruit N=40 participants and outline a control condition where individuals go through a standard conflict resolution training procedure. In the experimental condition, participants are additionally provided access to \tool{}, alongside the standard training procedure. In this section, we detail components of our user study and highlight aspects of end-to-end use across study participants. Our study aims to evaluate the following two hypotheses, centered around knowledge and application. 

\begin{quote}
    \textsc{\textbf{Knowledge Hypothesis}}: On a quiz, participants in the \tool{} condition will be more likely to recognize and/or recall specific strategies in the Interests-Rights-Power framework. 
\end{quote}

\begin{quote}
    \textsc{\textbf{Performance Hypothesis}}: In an applied setting with an evaluator (confederate) roleplaying conflict, participants in the \tool{} condition will be less likely to use contentious strategies (Rights, Power), and more likely to use cooperative strategies (Interests, Proposal, etc.).
\end{quote}

Our study finds little effect on ``book'' knowledge of IRP but strong effects on conflict performance, with participants roughly doubling their use of cooperative strategies and reducing the use of competitive strategies by 67\% in a live conflict. In this section, we outline our evaluation procedure and highlight the results for both hypotheses.

\subsection{Evaluation Conditions}
\label{eval_cond}
Our evaluation consists of two conditions: the control group, where participants see the status-quo of conflict resolution training, and \tool{}, where participants are additionally provided with our system to interactively practice conflict resolution. In each condition, participants are allocated \rev{a maximum of} 25 minutes of total practice, determined through a series of pilot studies where participants' completion times were measured.

\begin{itemize}
    \item \textbf{Control}: In the control condition, participants are provided a representative status quo of conflict resolution training: a tutorial video covering the IRP framework, and a list of conflict resolution strategies. No publicly available video covered all conflict resolution strategies, and existing slides were likewise inadequate for an introductory lecture. So, we recorded a 4-minute video covering all relevant conflict resolution strategies, modifying lecture slides from a pre-existing conflict resolution course at our institution. Additionally, participants can refer to a summary of the training video, with definitions and examples of conflict resolution strategies (similar to Table \ref{tab:conflict_strategies}).
    
    \item \textbf{\tool{}}: In the \tool{} condition, in addition to the training materials from the control condition, participants are additionally given access to \tool{}, and can practice conflict with simulations. Because participants are time-constrained to \rev{a max of} 25 minutes of practice, we randomly sample 3 scenarios from our full curated set of 12 (sample in Appendix \ref{case_studies}). Each participant was able to practice with the same 3 scenarios. 
\end{itemize}

\subsection{Procedure and Analysis}
\label{user_eval_procedure}
We conducted a between-subjects study of $N=40$ participants, measuring the causal effect of \tool{} as a system for teaching conflict. An overview of our study design is in Figure \ref{fig:study_design}. The evaluation consists of three components: a conflict \textbf{self-efficacy} quiz, a \textbf{knowledge} quiz that tests retrieval and recall of conflict resolution strategies, and a final roleplay that tests \textbf{application} skills.

All participants begin and end the study by completing a conflict self-efficacy survey. We administer the Dutch test, a set of 5-point Likert scales used to measure conflict self-efficacy. The test consists of 20 total scales, measuring five conflict resolution dimensions (compromising, problem-solving, yielding, forcing, avoiding) and producing a score from 4-20 for each dimension~\cite{de2001theory}. Results from the Dutch test have been verified across a range of conflict management studies~\cite{gelfand2012conflict, de2001minority, de2001managing}. Analogous to the Interests-Rights-Power strategies, ideal conflict results in high problem-solving/compromising scores, and low yielding/avoiding/forcing scores. The Dutch test also enables us to compute pre-post self-efficacy scores after participants complete the study. Because questions from each dimension are independent (no overlap), we conduct a paired $t$-test for the questions in each dimension, identifying if self-efficacy scores change within conditions following training. To identify differences across conditions, we fit a linear regression model for each dimension, predicting the post-efficacy scores to the pre-efficacy scores while treating the experimental condition as a dummy variable. We examine coefficients and $p$-values associated with the dummy condition variable.

For the \textbf{knowledge quiz}, we curate a set of 10 context messages (e.g. ``How about we spend more time working on the scheduling process instead?''), each relying on a single strategy from the full IRP framework. \rev{To keep the quiz short, we focus on core IRP strategies, choosing to omit \textit{Procedural} and \textit{Concession} as a result.} The full quiz is in Appendix \ref{knowledge_quiz_questions}; \rev{each participant receives the same quiz}. We ask participants to either recall or recognize the strategy used in each message. Recall questions are answered by providing the exact name of the strategy, while recognition questions require users to select the strategy from a multiple-choice form. The quiz first tests recall and then recognition---we do not allow participants to go back and change their answers, preventing use of the recognition options to answer recall questions. Additionally, we limit time allocated to both recall and recognition. Participants are given 3 minutes to answer all questions across both conditions. As with studying time, quiz time was determined through pilot studies. We conduct a two-sample $t$-test to across recall/recognition questions for both the \tool{} and control conditions. 

On finishing the knowledge quiz, all participants move to the \textbf{application} component of the evaluation. To evaluate a participant's ability to effectively apply conflict resolution strategies, we randomly select a scenario from \tool{}'s collected conflict training role-plays (discussed in \S\ref{premise_interaction}): participants never see this scenario before the evaluation (the withheld scenario is detailed in Appendix \ref{withheld_conflict}). Participants are then moved to a 1:1 chatroom with the evaluator. \rev{Consistent with our technical evaluation in \S\ref{tech_eval}, the evaluator is an author with significant exposure to conflict resolution training,} who is blinded to the participant's randomized training condition. \rev{All} participants then engage in a final conflict of 10 dialogue turns on the \rev{same withheld scenario}, for a total of 20 messages. Participants were both instructed to resolve the conflict and were informed that they were engaging with a real person. The evaluator carefully followed conflict spiral findings from \citet{brett1998breaking}, starting their roleplay with a contentious strategy, retaliating if contentious strategies were used, and yielding when cooperative strategies were repeatedly used. To ensure that the evaluator was consistent across conditions, the first and second author coded all messages with their primary IRP strategy (yielding a Cohen's Kappa~\cite{cohen1960coefficient} of 0.74), and tested for distributional differences between the conditional reply strategies $s$ used in each condition. Specifically, we used the two-sample Kolmogorov–Smirnov~\cite{massey1951kolmogorov} test across $P_{\tool{}}(reply_{s} | message_{s})$ and $P_{\textsc{Control}}(reply_{s} | message_{s})$, finding no significant difference between distributions for conflict resolution strategies ($p > 0.1$). In addition to being blinded to the condition, the evaluator behaved similarly in both conditions. After aggregating strategy use across the annotated conversations, we conduct 2-sample $t$-tests comparing each strategy's use across each condition. Following the application component of the study, participants retook the Dutch test. \rev{Finally, we} asked participants to leave optional open-ended comments on their training \rev{in a textbox shown at the end of the study.} 

Note that participants were offered no explicit incentive or motivation to actually utilize conflict resolution strategies during the study. Aside from an initial attention check early in the study, participants were free to spend as \rev{little} time as they wanted in \rev{training (e.g. using \tool{} and/or studying static training material) or conducting the final application component. Beyond the 25-minute maximum study time (\S \ref{eval_cond}), we enforce no other constraints.}

\subsection{Participant Details}

Participants for our study were recruited through Prolific, a crowdwork platform. We recruited participants who were over 18 years of age, lived in the U.S., and self-identified as fluent English speakers. Participants were also filtered to have at least 50 submissions and a 95\% approval rate, signed an IRB consent form approved by our institution, and were paid at a rate of \$12.00 per hour. 

\begin{table}[]
    \centering
    \begin{tabular}{l|rr}
        \toprule
        \textbf{Category} &\textbf{t-statistic} & \textbf{p-value}  \\
        \midrule
         \multicolumn{3}{c}{\textbf{\textsc{Performance Hypothesis}}} \\
        \midrule
        \multicolumn{3}{c}{\textit{Cooperative Strategies}} \\
        \midrule
        Interests & 3.87 & < 0.001*** \\
        Positive Expectations & 1.65 & 0.108\phantom{***} \\
        Proposal & 3.44 & < 0.001**\phantom{*} \\
        Concession & 1.04 &  0.304\phantom{***} \\
        \midrule 
        \multicolumn{3}{c}{\textit{Neutral Strategies}} \\
        \midrule
        Facts & -2.24 & 0.031*\phantom{**} \\
        Procedural Remarks & -1.04 & 0.304\phantom{***} \\
        \midrule         
        \multicolumn{3}{c}{\textit{Competitive Strategies}} \\
        \midrule
        Power & -2.29 & 0.028*\phantom{**} \\
        Rights & -2.09 & 0.043*\phantom{**} \\
        \midrule
         \multicolumn{3}{c}{\textbf{\textsc{Knowledge Hypothesis}}} \\
         \midrule
         Recall & 1.09 & 0.281\phantom{***} \\
         Recognition & 2.00 & 0.053\phantom{***}\\
         \bottomrule
    \end{tabular}
    \caption{2-sample $t$-test results across our study hypotheses. For our \textbf{\textsc{Performance Hypothesis},} where we evaluate application skills, we find that participants use significantly more cooperative strategies, and significantly fewer competitive strategies. In contrast, we cannot claim that \tool{} improves knowledge skills, failing to support our \textbf{\textsc{Knowledge Hypothesis}}.}
    \label{tab:assesment_stats}
\end{table}

\subsection{Results}

\begin{figure*}
    \centering
    \includegraphics[width=0.8\linewidth]{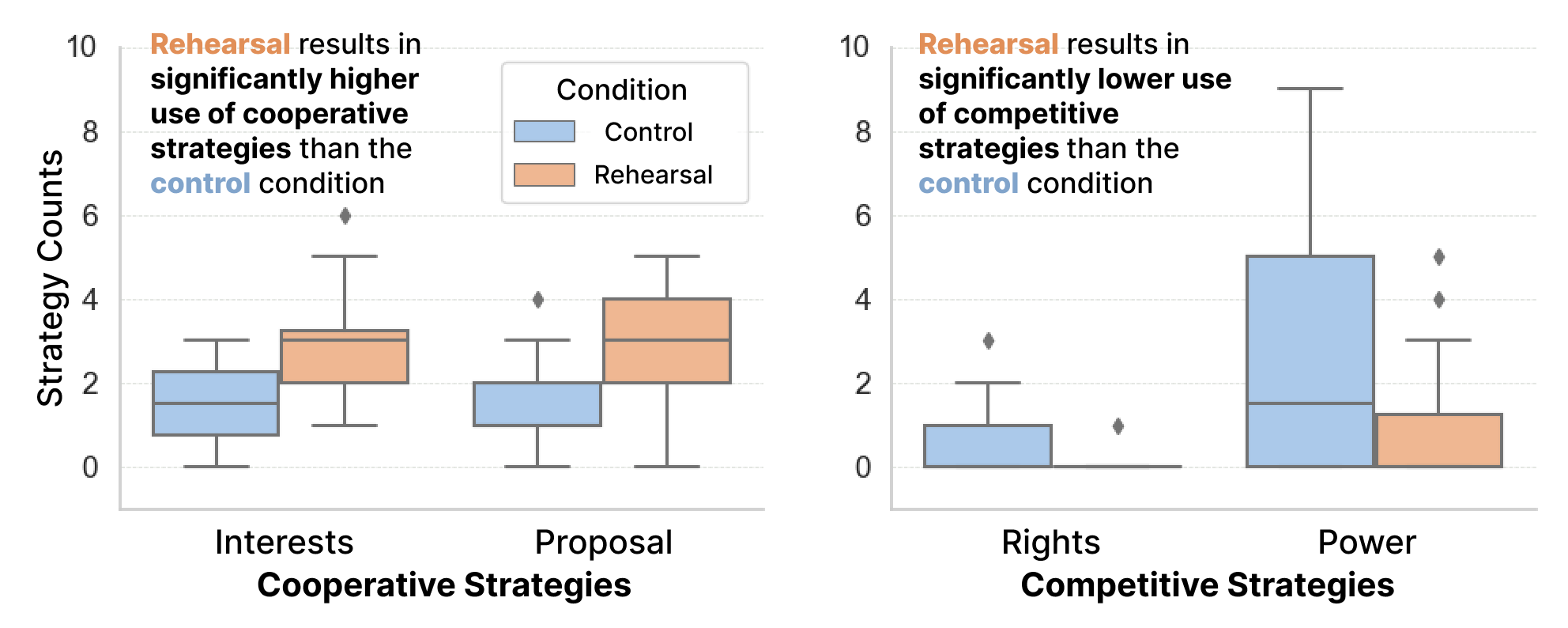}
    \caption{Cooperative and competitive strategy use between both the \tool{} and control conditions in our user study.}
    \label{fig:strat_counts} 
\end{figure*}

\paragraph{Practice with \tool{} results in increased use of cooperative strategies and decreased use of competitive strategies.} Using \tool{} has a clear effect on conflict strategy use in the application portion of our evaluation. In the final 10-turn chat, participants using \tool{} were twice as likely to use the Interests and Proposal-oriented strategies compared to the control condition, averaging 3.0 vs. 1.5 times for Interests, and 2.9 vs. 1.5 times for Proposal. Both strategies show significant improvement compared to the control condition (2-sample $t$-test, $p$ < 0.05). Furthermore, participants in \tool{} reduce their use of competitive strategies. Use of Power strategies in the \tool{} condition is 1/3 of the control condition (2.9 vs 1.0). Rights-based strategies are almost never used in the \tool{} condition ($\approx$ 0) but appear on average 0.5 times per conversation in the control condition. Differences across conditions for both the Power and Rights strategies are also significant ($p< 0.05$). Participants with \tool{} are far more performant in the actual conflict, supporting the \textsc{\textbf{Performance Hypothesis}}.

\paragraph{Passively reading or watching conflict resolution training material builds skills orthogonal to actual practice} Results from the application component of our evaluation are universally positive: participants in \tool{} significantly outperform participants in the control condition. Participants from the control condition offer a simple explanation for our quantitative results:

\begin{quote}
    ``It's easy to read about conflict-resolution strategies, but a lot harder to implement and stick to.''
\end{quote}
Simply reading over strategies does not translate over to applying them. Participants mistakenly feel prepared after watching/reading training material and are blindsided by the difficulty of the application component in the evaluation.
\begin{quote}
    ``I'll be honest: while I did pay attention to the training, it was really hard to imagine implementing [the conflict resolution strategies] without feeling like a total ass. I feel like it's obvious when people read about some weirdo social technique or something and try to do it in conversation, they end up sounding like brainwashed robots.''
\end{quote}

A common challenge across participants in the control condition lies in application. \tool{} shines at providing application-oriented feedback to participants: as we've observed, application is primarily where we see outsized effects in our final evaluation. While it's challenging for participants in the control condition to ``imagine implementing'' conflict resolution strategies, \textit{\tool{} offers exactly this!} Practicing implementation using a simulated roleplay ensures that participants in the \tool{} condition are more prepared to apply and implement strategies in the evaluation.

\paragraph{While \tool{} builds application skills, simulated practice does not significantly improve recognition or recall} Across both conditions, the average quiz score is 12.6/20 or 63\%. When stratifying by experimental condition, we see a difference in performance: 11.3/20 for control vs. 13.8/20 for \tool{}. Using \tool{}, participants also see an average overall increase in recall/recognition quiz scores (5.3/10 $\rightarrow$ 6.2/10 for recall and 6.0/10 $\rightarrow$ 7.6/10 for recognition). However, these differences are not significant, with our $t-$test indicating no effect ($p > 0.05$). While tool helps with application, we find no support for the \textsc{\textbf{Knowledge Hypothesis}}. This result suggests that \tool{} improves ``street knowledge'' but not necessarily ``book knowledge'' of conflict---which, if we had to choose between the two types of knowledge, is the preferable outcome. 

\begin{figure}
  \begin{center}
    \includegraphics[width=0.3\textwidth]{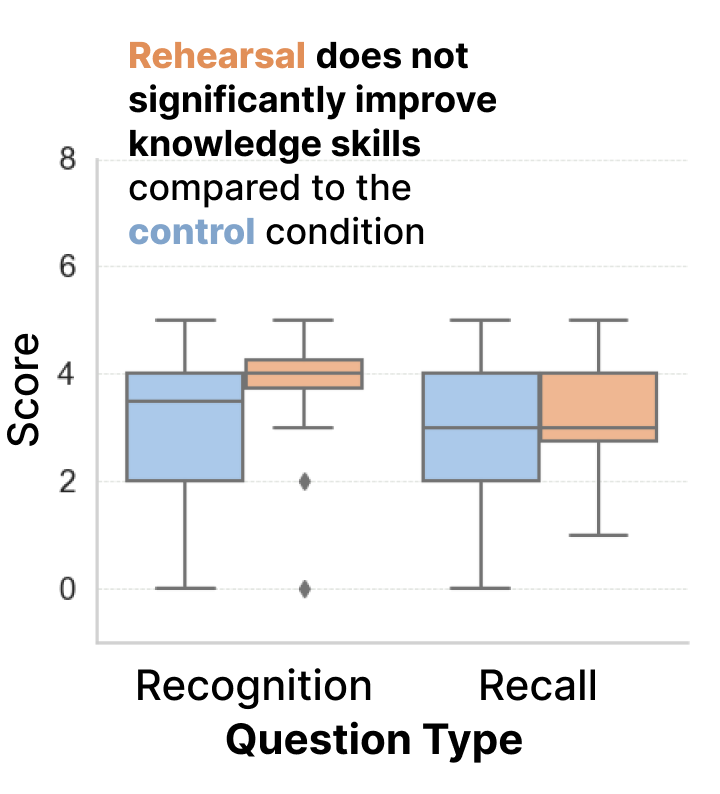}
  \end{center}
  \caption{Knowlege quiz scores stratified across recall/recognition. }
  \label{fig:knowledge}
\end{figure}

\paragraph{The ``Conflict Reality Check:'' participants reduce conflict self-efficacy after training} Participants in both conditions generally \textit{reduced} their Dutch conflict self-efficacy scores following the final conflict in the study. Participants in both conditions indicate that they use fewer cooperative strategies (compromising, yielding, and problem-solving) strategies, but use more competitive strategies, specifically forcing (paired $t$-test, $p < 0.05$). Furthermore, results from our regression model indicate that participants in the \tool{} condition self-reported \textit{increased} use of forcing and \textit{decreased} use of yielding compared to the control group ($p < 0.05$).

This is at odds with our empirical application results. Firstly, any practice should increase efficacy scores, but we observe a decrease for all cooperative dimensions. Secondly, participants in the \tool{} condition clearly use fewer competitive strategies, but self-report significantly higher use of forcing and lower use of yielding strategies. What gives? We suspect that is a classic example of the Dunning Krueger effect~\cite{kruger1999unskilled}. Participants after any training have a better understanding of their true conflict resolution skills, and recalibrate their scores---moving them down!---based on their performance in the evaluation. In the \tool{} condition, participants are even better at recognizing power-oriented forcing strategies, and reduce their scores to a larger degree than the control condition.

\paragraph{You can lead a participant to water...} In aggregate, participants using \tool{} significantly outperformed participants in the control condition during the actual conflict. However, even in the \tool{} condition, 2 participants employed just a \textit{single} interests-oriented strategy in the final evaluation conflict. Surprisingly, both participants performed fairly well on the quiz component of the study, averaging 7.5/10 correct answers. One of these participants left the following comment:

\begin{quote}
    ``I felt like the training helped me learn new methods, but I am stubborn by nature and once the person was rude from the outset, I did not want them to get their way.''
\end{quote}
These cases highlight the scope of \tool{}'s purpose. While \tool{} improves use of conflict strategies on average, it offers no guarantee that a specific participant will actually choose to use them. An individual who comes in with no intent to actually resolve a conflict likely will not change their mindset following any kind of training. 

\subsection{\tool{}'s Failure Modes}
A two-stage inductive coding of \tool{}'s interaction traces across participants highlights a range of failure modes, some more egregious than others. Anomalies in individual message generations were coded and repeating themes were merged. We describe these anomalies below:  

\paragraph{Occasionally, simulations feel overly scripted.} We note that some simulated messages are too ``on the nose'': since \tool{} generates messages conditioned on a strategy. For example, \textit{Rights} generated messages explicitly mention ``rights'' or ``fairness'' (e.g. ``That's not fair!''); \textit{Proposal} messages explicitly mention a proposal (e.g. ``I propose we do this...''). In realistic conflict, the recognition and application of strategies are more nuanced. \textit{Rights}-based strategies depend on the context of a conflict and the norms that shape it. What constitutes a \textit{Power} strategy depends on complex power dynamics between individuals in a party. While \tool{} allows users to specify properties of their own conflict in custom scenarios, we do not rigorously evaluate this feature in our user studies.  

\paragraph{Hallucinations, though mitigated by IRP, still appear.} LLMs often make up facts about a scenario to justify an argument (e.g. disagreeing about the specifics of being late to work). For most scenarios, this is acceptable. In roleplay, experts will introduce facts as a conflict progresses. However, we have limited control over when LLMs hallucinate, which may detract from practicing the intended scenario. Related work on story generation~\cite{yang-etal-2022-re3} proposes the use of an edit module to store and correct factual inconsistencies. Integrating this into IRP prompting is an avenue for future work.

\paragraph{For a specific scenario, generated conversations are less diverse than they initially seem.} For our evaluation, participants spend a limited amount of time practicing with \tool{}, engaging in a few conversations across 3 scenarios. Given this limited exposure, we suspect that participants themselves did not notice diversity problems. Zooming out across all interactions, however, we note that the style of conflict is fairly regular within a scenario. Simulated conversations follow a prescriptive and formal style. We suspect that instruction-following in LLMs serves as a strong prior over the style of generated text. Manually finetuning open-source LMs to simulate conflict will likely mitigate this issue.

\section{Discussion}
In this section, we reflect on teaching through strategic simulation and grounding LLMs in social science theories. We also discuss ethical risks related to applying generative models to educational applications.

\subsection{Simulation as an Effective Teaching Interaction}

Roleplay has long been an established method for teaching across a wide range of domains: customer service~\cite{borner2012staff}, language learning~\cite{livingstone1983role}, therapy~\cite{lazarus1966behaviour}, healthcare~\cite{nestel2007role}, and more. Across these domains, effective roleplay requires an expert to simulate a specific scenario, while working meticulously through a domain-specific teaching process. \tool{}'s goal is to focus on a setting where roleplay serves a foundational role---conflict resolution---and highlight how effective roleplay simulation can yield strong results. More broadly, we argue through this work that applying simulated roleplay requires understanding when, and for what skills, simulation is effective.

\paragraph{When does simulated roleplay work?} We hypothesize that simulated roleplay has a higher likelihood of working in domains where expert roleplay is already effective. In these domains, experts have already laid the groundwork for roleplay---frameworks like IRP exist \textit{because} experts regularly teach conflict resolution. Similarly, \tool{}'s effectiveness is only possible because of its dependence on these frameworks. Remaining faithful to social scientific theory that already works in practice---and ensuring that LLM generations are constrained to a selected theory---will increase the chances of successfully simulating roleplay in those domains.  IRP prompting, for example, depends heavily on how experts employ conflict resolution strategies to simulate conflict. Prior work in roleplay lies primarily in soft-skill training, so we expect successful applications to operate primarily in these contexts.

\paragraph{What kinds of skills does simulated roleplay build?} After evaluating \tool{}, we find a significant increase in participants using cooperative conflict resolution strategies compared to a control group. However, when asked to recall and recognize strategies, participants show \textit{no significant difference} between groups. \tool{}'s primary interaction revolves broadly around applying cooperative and competitive strategies. There is little incentive for users to build ``book smarts'', i.e. the specific names associated with each of the 8 IRP strategies. For example, our Recall and Recognition interaction---where we ask users to recall a simulation's strategy---occupies a small part of the full interaction loop. All that matters is the ``street smarts'', or the application of these principles. Given this incentive misalignment, users in a simulated roleplay likely prioritize application over memorization, as reflected in our results. Therefore, while simulations like \tool{} can be beneficial for the practical application of skills such as conflict resolution, they may not significantly improve specific knowledge skills. In domains where learning these skills is critical, educators can separately apply both experiential learning through simulations and traditional educational methods (e.g. flash cards, spaced repetition).

\subsection{Grounding in Theory \textit{Improves} LLM Generation Capabilities}
\tool{} works by constraining the generation of LLMs to a specific theory (in our case, IRP). Constrained generation is already an established trend in a range of more traditional NLP tasks. From code generation~\cite{yin2017syntactic} to summarization to story generation~\cite{yang-etal-2022-re3, yang-etal-2023-doc}, LLMs augmented with explicit planning/constraining components generally outperform out-of-the-box models~\cite{gandhi2023strategic}. Our work, however, takes traditional constrained generation a step further, using internal social-scientific knowledge of an LLM as guardrails for the same LLM's generation process. More concretely, we ground LLMs in an established subset of social scientific theory, augmenting models with latent, specialized knowledge of interpersonal interaction. Given that large language models are trained on vast swaths of the internet, we suspect that they contain information relevant to a range of established social scientific theories. IRP prompting first elicits this knowledge (with a planning component), then ties LLM generation to the elicited knowledge. 

Across our evaluations, we find that grounding to IRP theory improves the validity of \tool{} as a teaching tool. Instead of sampling randomly across a larger space of generations, constraining an LLM specifically to a teachable theoretical framework significantly increases the effectiveness of \tool{}. Users have an actionable and controlled space to practice; generations are ``predictable enough,'' providing users with a practice sweet spot. While unconstrained generation may offer more variety, a lack of control during generation yields less effective simulations. 

Still, constraining an LLM with its own internal knowledge is not a straightforward feat. A core challenge of building \tool{}---and more specifically, IRP prompting---lies in eliciting and evaluating this knowledge, and applying it to a carefully constrained task. The IRP prompting pipeline is a multi-step process, with each step supervising a model's final output. Even with multi-step prompts, we conduct careful evaluations of an LLMs ability to use IRP effectively, ablating components in our pipeline and evaluating an LLMs ability to classify conflict resolution strategies. 

In theory, one could simply substitute the IRP definitions in our prompts with their own framework, and allow our planning components to elicit knowledge and constrain generation. Indeed, we suspect that general patterns from IRP prompting can be applied to different social scientific theories, enabling a \tool{}-like teaching interactions for a wider range of domains. While grounding in IRP proves successful for \tool{}, there is no guarantee that LLMs possess adequate latent knowledge across other theoretical frameworks. For novel social scientific theories not already present in the training data, we are unsure of our pipeline's effectiveness. Successful constrained generation requires that LLMs can apply and recognize the target theory first.

\subsection{Limitations and Future Work}
While \tool{} proves effective in our user study, there are important limitations and avenues for future work. 

\paragraph{IRP does \textbf{not} cover all conflict resolution scenarios} Conflict resolution theory assumes that both parties can resolve conflict in good faith. This is definitely not true of all conflict: there are instances where no application of cooperative strategies will cause the interlocutor to change their position.\footnote{Every researcher has at least one such story about an intransigent reviewer.}
Conflict scenarios that might be actively dangerous to engage in, or pose too high of a risk to try resolving, may require additional support or even disengagement. \tool{} cannot extrapolate to situations like these. Future work should predict if a specific conflict falls under IRP's scope, and warn against the effectiveness of simulated practice under these settings. Finally, IRP is a Eurocentric conflict resolution framework. Strategies vary significantly across cultures and should be an important consideration for future work~\cite{gelfand2012conflict}.

\paragraph{Near and Far Transfer} We evaluate performance on conflict resolution tasks immediately after simulated training. Therefore, we can only claim that \tool{} is effective with \textit{near transfer} in learning, where skills are retained immediately after an educational intervention~\cite{perkins1992transfer}. We do not test if \tool{}'s effect persists over a longer period of time via \textit{far transfer}~\cite{barnett2002and}. Evaluating \textit{far transfer} is an important avenue for future work. 

\rev{\paragraph{Additional Applications} \tool{} currently supports two parties during a conflict. Conflict, however, often involves multiple parties. Future work can extend \tool{} to multiparty setups, where each party is powered by IRP prompting. Supporting multiparty conflict resolution enables training across more general moderation and consensus-building activities. Finally, as generative models continue improving in domains beyond text, we also expect training systems like \tool{} to support multimodal interaction (e.g. voice, video).}

\subsection{Ethical and Societal Considerations}
While simulated roleplay shows promise as a useful teaching interaction, we must also consider critical ethical and societal considerations of deploying systems like \tool{}.

\paragraph{Deployment Risks and Distributional Shift}
\tool{} is designed primarily as a safe training environment in which users can practice their skills and learn how to transfer those skills to the real world. As with other training environments, we acknowledge this shift between training and the real world. \tool{} is only intended for simulation; it may not reflect reality exactly. While practicing with simulated roleplays, we urge users to use \tool{} with caution and to keep in mind these risks.

\paragraph{Stereotypes} LLMs have been documented to output a range of stereotypes~\cite{bolukbasi2016man, feng2023pretraining}. Therefore, \tool{} might be likely to reflect both negative and positive stereotypes, especially since we rely on personas in each scenario~\cite{cheng-etal-2023-marked} and use a multi-step prompting strategy~\cite{shaikh2022second}. To mitigate this risk,  deployments of \tool{} should explicitly highlight important attributes (e.g., "male", "aggressive", and "ambitious") in the prompts and enable users to make changes as needed in the interface. This approach explicitly reminds users of potential stereotypes associated with their prompts, and gives users full control over the types of personas they wish to communicate with; this allows for more direct engagement between users and the tool. Furthermore, \tool{} is likely to follow any descriptions given explicitly in the prompt, so stereotypes are also likely to arise from under-description in the prompt (e.g., if the prompt says that the interlocutor is just a "boss", the model will likely stereotype them as a white male). Deployments of \tool{} must explicitly spell out any characteristics that the model should be operationalizing, limiting its ability to stereotype.

\paragraph{Job Displacement} A final risk is that \tool{} may lead to job displacement or devaluation for expert trainers. However, this would require that people who currently pay professional trainers for conflict training stop doing so. Such trainers are generally retained by extremely wealthy firms, and it seems unlikely that such firms would stop doing so. More likely, such trainers would integrate these tools as part of their education and training events. Similarly, chemistry simulation tools don't replace chemistry classes, and management books don't replace management coaching. So, we expect that the high end will likely remain stable. It is certainly possible that some individuals on the margin will opt for a cheap or free standalone option. We place this risk against the potential benefit of a free-to-use tool, which can benefit a broader user population, especially those without expensive professional training or social capital. It will be easier for professional experts to handle a larger population and to focus on more tailored, challenging scenarios needed by skill training, and they will still maintain their high level of expertise, maintaining their uniqueness.

\section{Conclusion}
We introduced \tool{}, a system for teaching conflict resolution by simulating conflict. While using \tool{}, individuals interacted with a simulated interlocutor, exercising their conflict-resolution skills. \tool{} is powered by IRP prompting, a multi-step prompting technique that grounds language models in conflict resolution theory. Through interaction with grounded simulations of conflict, users built an intuition for applying effective conflict resolution strategies. We conducted a between-subjects evaluation of $N=40$ participants, who engaged in an actual conflict following training with or without the tool. Participants with \tool{} training significantly improved their application of effective conflict resolution strategies in the unaided conflict. Finally, we discussed applications of \tool{} beyond conflict resolution; and highlighted limitations, ethical considerations, and societal impacts. 


\begin{acks}
We thank Jayanthi Subramanian, Jordan Troutman, Jensen Gao, Lindsay Popowski, Michelle Lam, Tiziano Piccardi, Shan Rizvi, Jiaju Ma, Yilu Sun, and Ben Prystawski for insightful discussion and feedback while designing and evaluating \tool{}. We additionally thank Nathan Fielder for his show, The Rehearsal, which inspired \tool{}'s name. Omar Shaikh was supported by the Brown Institute's Magic Grant. Additionally, our work was supported by Google, NSF Award CCF-191894, NSF IIS-2247357, and the Stanford Institute for Human-Centered Artificial Intelligence (HAI).
\end{acks}

\bibliographystyle{ACM-Reference-Format}
\bibliography{sample-base}


\appendix

\section{Conflict Case Studies}
\label{case_studies}

\label{premises}
Here, we detail a subset of premises from the Harvard Program on Negotiation~\cite{ponharvard} and the Crucial Conversations book~\cite{crucial_learning} that are used by \tool{} during evaluation. 

\subsection{Practice Case Studies}
\paragraph{Undercooked meal.} 

\begin{quote}
``You just tried a meal your partner cooked for you, but it's slightly undercooked. You mention this to your partner, and they're visibly unhappy that you brought this up.''   
\end{quote}

\paragraph{Where's my refund?} 
\begin{quote}
``The complaints clerk (you) in a department store sees a customer (Casey) coming with a blender. 
The store cannot return these items to the manufacturer. 
You have a small weekly budget to absorb the cost of such items, if returned, and the department head has instructed that it be used sparingly. The budget for this week is overspent. Casey, having used the blender for over a week, believes it is either defective or an inadequate appliance, and has therefore decided to return it, and is angrily demanding a refund.'' 
\end{quote}

\paragraph{Work Performance}
\begin{quote}
``Jerry has been a steady employee for four years. 
Recently, Jerry's work and attitude have taken a turn for the worse. 
Jerry's supervisor (Casey) does not know why, but the situation has come to the point where the supervisor is prepared to fire Jerry, and is under considerable pressure from management to do so. 
The two are about to meet to discuss this situation.''
\end{quote}

\subsection{Withheld Evaluation Case Study}
\label{withheld_conflict}
\paragraph{The Unwanted Promotion.}
\begin{quote}
    ``Your boss Chris keeps telling
    you that you’d make a great supervisor. You
    don’t want the promotion. You like what you
    do. Chris said team players take
    promotions. You’ve heard that Chris
    is submitting the paperwork to have you
    promoted. Yesterday Chris said you’d soon be
    getting a big surprise. This morning he asked
    you to be sure to go to the afternoon team
    meeting. You don’t want him to spring the
    announcement in the meeting and pressure
    you. You're now in a 1:1 meeting with him, and he's annoyed that you're planning on turning this down.''
\end{quote}

\section{Knowledge Quiz}
\label{knowledge_quiz_questions}
Our Knowledge quiz is split into two categories, with 5 questions per category (recall and recognition)

\subsection{Recall Questions}
\begin{enumerate}
    \item I'm going to have to report you to your manager. \textbf{Power}
    \item How about we spend more time working on the scheduling process instead? \textbf{Proposal}
    \item I totally understand where you're coming from. \textbf{Interests}
    \item If we work together, I'm sure we can figure out what's wrong. \textbf{Positive Expectations}
    \item I think you're breaking company policy here... \textbf{Rights}
\end{enumerate}

\subsection{Recognition Questions}
\begin{enumerate}
    \item I'll destroy your career if you come in here complaining again. \textbf{Power}
    \item I put in 60 hours for the last 4 weeks. \textbf{Facts}
    \item I really wanted a promotion this year. \textbf{Interests}
    \item Didn't we agree to this? This is so unfair. \textbf{Rights}
    \item I can get that to you tommorow. How does that sound? \textbf{Proposal}
\end{enumerate}


\end{document}